%% file: ms_tk_cr.tex
\documentclass[useAMS,usenatbib]{mn2e}
\usepackage{color}
\usepackage{array} 
\usepackage{amsmath}
\usepackage{graphicx}
\usepackage{epstopdf}
\usepackage[labelfont=bf,labelsep=period,justification=raggedright,singlelinecheck=false,font=small]{caption}
\usepackage{wrapfig}
\usepackage{float}
\usepackage{amssymb}
\usepackage[normalem]{ulem}
\usepackage{hyperref}
\usepackage{undertilde}
\usepackage[T1]{fontenc}
\usepackage[utf8]{inputenc}
\usepackage{comment}
\usepackage[export]{adjustbox}
\usepackage{scalerel}
\usepackage{tikz}
\usetikzlibrary{svg.path}

\usepackage{ctable}

\definecolor{orcidlogocol}{HTML}{A6CE39}
\tikzset{
  orcidlogo/.pic={
    \fill[orcidlogocol] svg{M256,128c0,70.7-57.3,128-128,128C57.3,256,0,198.7,0,128C0,57.3,57.3,0,128,0C198.7,0,256,57.3,256,128z};
    \fill[white] svg{M86.3,186.2H70.9V79.1h15.4v48.4V186.2z}
                 svg{M108.9,79.1h41.6c39.6,0,57,28.3,57,53.6c0,27.5-21.5,53.6-56.8,53.6h-41.8V79.1z M124.3,172.4h24.5c34.9,0,42.9-26.5,42.9-39.7c0-21.5-13.7-39.7-43.7-39.7h-23.7V172.4z}
                 svg{M88.7,56.8c0,5.5-4.5,10.1-10.1,10.1c-5.6,0-10.1-4.6-10.1-10.1c0-5.6,4.5-10.1,10.1-10.1C84.2,46.7,88.7,51.3,88.7,56.8z};
  }
}

\newcommand\orcidicon[1]{\href{https://orcid.org/#1}{\mbox{\scalerel*{
\begin{tikzpicture}[yscale=-1,transform shape]
\pic{orcidlogo};
\end{tikzpicture}
}{|}}}}

\def\apj{ApJ}
\def\aap{A\& A}
\def\apjl{ApJL}
\def\apjs{ApJS}
\def\mnras{MNRAS}

\def\pasa{PASA}
\def\aj{AJ}
\def\araa{ARA\& A}
\def\nat{Nature}
\def\prd{Phys. Rev. D}

\def\rmxaa{RMxAA}

\def\aapr{A\&A~Rev.}
\def\pasj{PASJ} 
\def\apss{Ap\&SS}
\def\prl{Phys. Rev. Lett.}

\newcommand\altaffilmark[1]{$^{#1}$}
\newcommand\altaffiltext[1]{$^{#1}$}

\newcommand{\msun}{{\rm M}_{\odot}}

\restylefloat{table}

\begin{document}

\title[CRs in disk-halo interface]{The impact of cosmic rays on dynamical balance and disk-halo interaction in $L\star$ disk galaxies}

\author[T. K. Chan et al.] {T. K. ~Chan\altaffilmark{1,2} \thanks{Email: (TKC) tsang.k.chan@durham.ac.uk}$^{\orcidicon{0000-0003-2544-054X}}$,
  Du\v{s}an ~Kere\v{s}\altaffilmark{2} \thanks{Email: (DK) dkeres@physics.ucsd.edu} $^{\orcidicon{0000-0002-1666-7067}}$,
  Alexander B. ~Gurvich\altaffilmark{3}$^{\orcidicon{0000-0002-6145-3674}}$,
  Philip F. ~Hopkins\altaffilmark{4}$^{\orcidicon{0000-0003-3729-1684}}$,\newauthor
  Cameron ~Trapp\altaffilmark{2}$^{\orcidicon{0000-0001-7813-0268}}$,
  Suoqing ~Ji\altaffilmark{4,5}$^{\orcidicon{0000-0001-9658-0588}}$,
  Claude-Andr{\'e} ~Faucher-Gigu{\`e}re\altaffilmark{3}$^{\orcidicon{0000-0002-4900-6628}}$
 \vspace*{6pt}
  \\
  \altaffiltext{1}{ Institute for Computational Cosmology, Department of Physics, Durham University, South Road, Durham DH1 3LE, UK}\\
  \altaffiltext{2}{Department of Physics, Center for Astrophysics and Space Sciences,University of California San Diego,}\\ {9500 Gilman Drive, La Jolla, CA 92093, USA}\\
  \altaffiltext{3}{Department of Physics \& Astronomy and CIERA, Northwestern University, 1800 Sherman Ave, Evanston, IL 60201, USA}\\  
  \altaffiltext{4}{TAPIR, Mailcode 350-17, California Institute of Technology, Pasadena, CA 91125, USA.}\\
  \altaffiltext{5}{Astrophysics Division \& Key Laboratory for Research in Galaxies and Cosmology, Shanghai Astronomical Observatory,}\\ {Chinese Academy of Sciences, 80 Nandan Road, Shanghai 200030, China.} \vspace*{6pt}
}

\maketitle

\begin{abstract}
Cosmic rays (CRs) are an important component in the interstellar medium (ISM), but their effect on the dynamics of the disk-halo interface (< 10 kpc from the disk) is still unclear. We study the influence of CRs on the gas above the disk with high-resolution FIRE-2 cosmological simulations of late-type $L\star$ galaxies at redshift $z\sim 0$. We compare runs with and without CR feedback (with constant anisotropic diffusion $\kappa_\parallel \sim 3\times10^{29}{\rm cm^2/s}$ and streaming). Our simulations capture the relevant disk halo interactions, including outflows, inflows, and galactic fountains. Extra-planar gas in all of the runs satisfies {\it dynamical balance}, where total pressure balances the weight of the overlying gas. While the kinetic pressure from non-uniform motion ($\gtrsim 1\;{\rm kpc}$ scale) dominates in the midplane, thermal and bulk pressures (or CR pressure if included) take over at large heights. 
 We find that with CR feedback, (1) the warm ($\sim 10^4~{\rm K}$) gas is slowly accelerated by CRs; (2) the hot ($> 5\times 10^5~{\rm K}$) gas scale height is suppressed; (3) the warm-hot ($2\times 10^4-5\times10^5~{\rm K}$) medium becomes the most volume-filling phase in the disk-halo interface. 
 We develop
 a novel conceptual model of the near-disk gas dynamics in low-redshift $L\star$ galaxies: 
 with CRs, the disk-halo interface is filled with CR-driven warm winds and hot super-bubbles that are propagating into the CGM with a small fraction falling back to the disk. 
Without CRs, most outflows from hot superbubbles are trapped by the existing hot halo and gravity, so typically they form galactic fountains. 
\end{abstract}
\begin{keywords}
ISM: cosmic rays --- galaxies: spiral --- galaxies: kinematics and dynamics --- ISM: kinematics and dynamics
\end{keywords}
\label{firstpage}
\input{introduction}
\input{method}
\input{results}

\input{discussions}

\input{conclusions}

\section*{ACKNOWLEDGEMENTS}
 We gratefully thank the referee, Mordecai-Mark Mac Low, for the careful reading and suggestions. We thank Bili Dong for the help with YT. We thank Eliot Quataert, Norman Murray, and Matthew Orr for helpful discussions. We are grateful for stimulating ideas and discussions during the KITP program -- Fundamentals of Gaseous Halos (grant number NSF PHY-1748958). TKC was supported by the Science and Technology Facilities Council (STFC) through Consolidated Grants ST/P000541/1 and ST/T000244/1 for Astronomy at Durham. 
DK was supported by NSF through grants: AST-1715101 and AST-2108314 and Cottrell Scholar Award from the Research Corporation for Science Advancement.
ABG was supported by a National Science Foundation Graduate Research Fellowship Program under grant DGE-1842165 and was additionally supported by the NSF under grants DGE-0948017 and DGE-145000, and from Blue Waters as a graduate fellow which is itself supported by the NSF (awards OCI-0725070 and ACI-1238993). Support for PFH and co-authors was provided by an Alfred P. Sloan
Research Fellowship, NSF Collaborative Research Grant \#1715847 and CAREER grant
\#1455342, and NASA grants NNX15AT06G, JPL 1589742, 17-ATP17-0214. SJ is supported by a Sherman Fairchild Fellowship from Caltech, the Natural Science
Foundation of China (grants 12133008, 12192220, and 12192223), and the science
research grants from the China Manned Space Project (No. CMS-CSST-2021-B02). CAFG was supported by NSF through grants AST-1715216, AST-2108230, and CAREER award AST-1652522; by NASA through grant 17-ATP17-0067; by STScI through grant HST-AR-16124.001-A; and by the Research Corporation for Science Advancement through a Cottrell Scholar Award. The simulation presented here used computational resources granted by the Extreme Science and Engineering Discovery Environment (XSEDE), which is supported by National Science Foundation grant no. OCI-1053575, specifically allocation TG-AST120025. The simulations are additional supported by allocations AST21010 and AST20016 from the NSF and TACC. Our analysis is run on the Triton Shared Computing Cluster in the San Diego Supercomputer Center \citep{TSCC}. This work also made use of {\small YT} \citep{YTproject}, matplotlib \citep{Hunt07matplotlib}, numpy \citep{vand11numpy}, scipy \citep{Jone01scipy}, and NASA’s Astrophysics Data System. The data used in this work here were, in part, hosted on facilities supported by the Scientific Computing Core at the Flatiron Institute, a division of the Simons Foundation.

This is a pre-copyedited, author-produced PDF of an article accepted for publication in Monthly Notices of the Royal Astronomical Society following peer review. The version of record is available online at: https://doi.org/10.1093/mnras/stac2236 .

\section*{DATA AVAILABILITY}
The data underlying this article will be shared on reasonable request to the corresponding author (TKC). The simulation initial conditions, snapshot files, and visualization can be found in \href{https://fire.northwestern.edu/data/}{https://fire.northwestern.edu/data/}. A public version of the GIZMO simulation code is available \href{http://www.tapir.caltech.edu/~phopkins/Site/GIZMO.html}{http://www.tapir.caltech.edu/~phopkins/Site/GIZMO.html}.

\bibliographystyle{mn2e}
\bibliography{mn-jour,mybib}

\appendix
\section{The effects of MHD and CR diffusion coefficient}
\label{sec:mhdcr28}

\begin{figure*}
 \includegraphics[width={0.95\textwidth}]{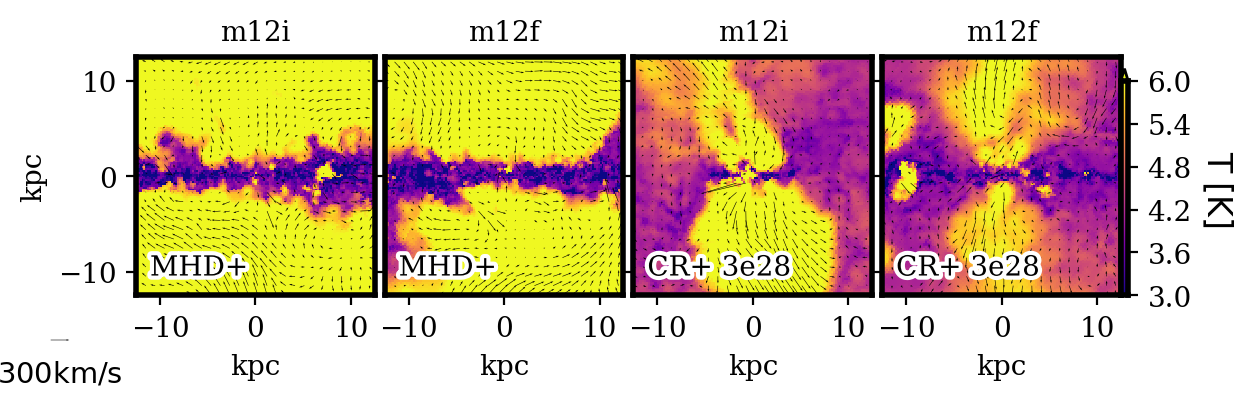}
\caption{Edge-on slides of gas temperature profiles, cutting through the galactic centers. Arrows represent gas velocities. MHD+ are similar to Hydro+, but CR+3e28 has more hot gas outflows than CR+(3e29) (see the main text).}
\label{fig:hydromhdcrquiver}
\end{figure*}

\begin{figure*}
\includegraphics[width={0.4\textwidth}]{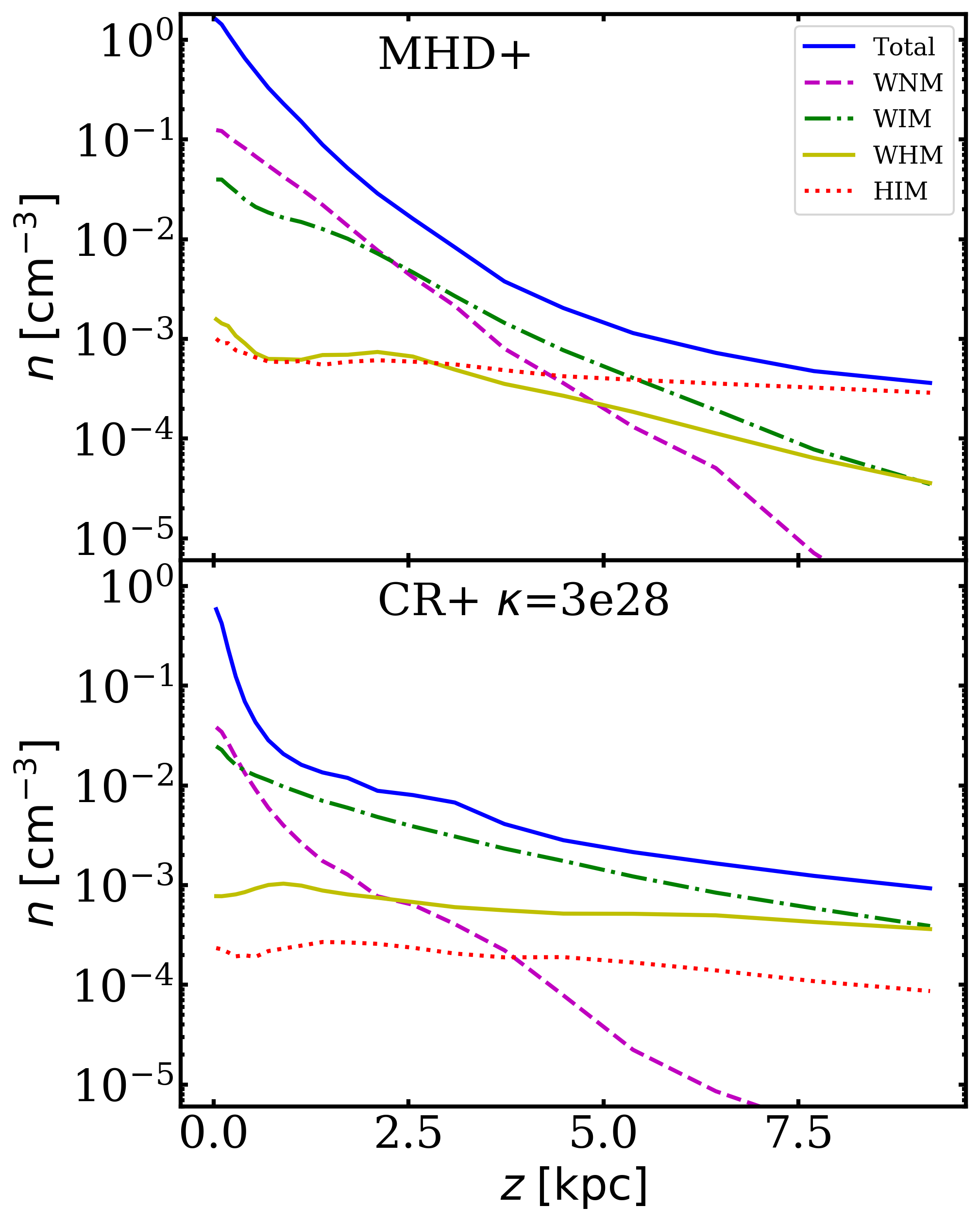}
  \includegraphics[width={0.4\textwidth}]{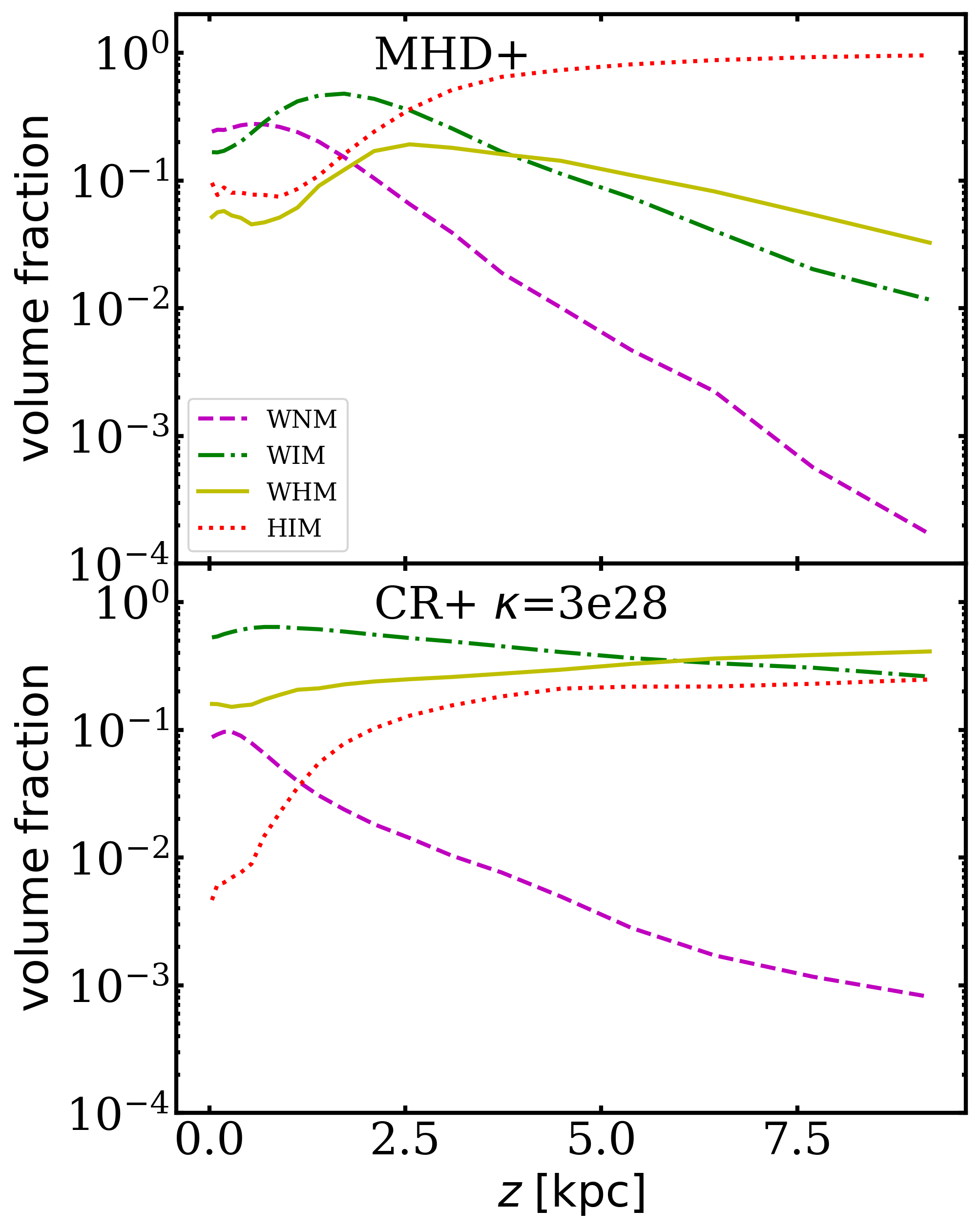}
\caption{ Vertical distributions of volume-weighted gas density (left) and volume filling factor (right) at different gas phases with $R<10\;{\rm kpc}$ in {\bf m12i} averaged over last 250 Myr (labels as in Fig. \ref{fig:gasdenTz}).}
\label{fig:gasdenTznc_m12crb70}
\end{figure*}

\begin{figure*}
 \includegraphics[width={0.75\textwidth}]{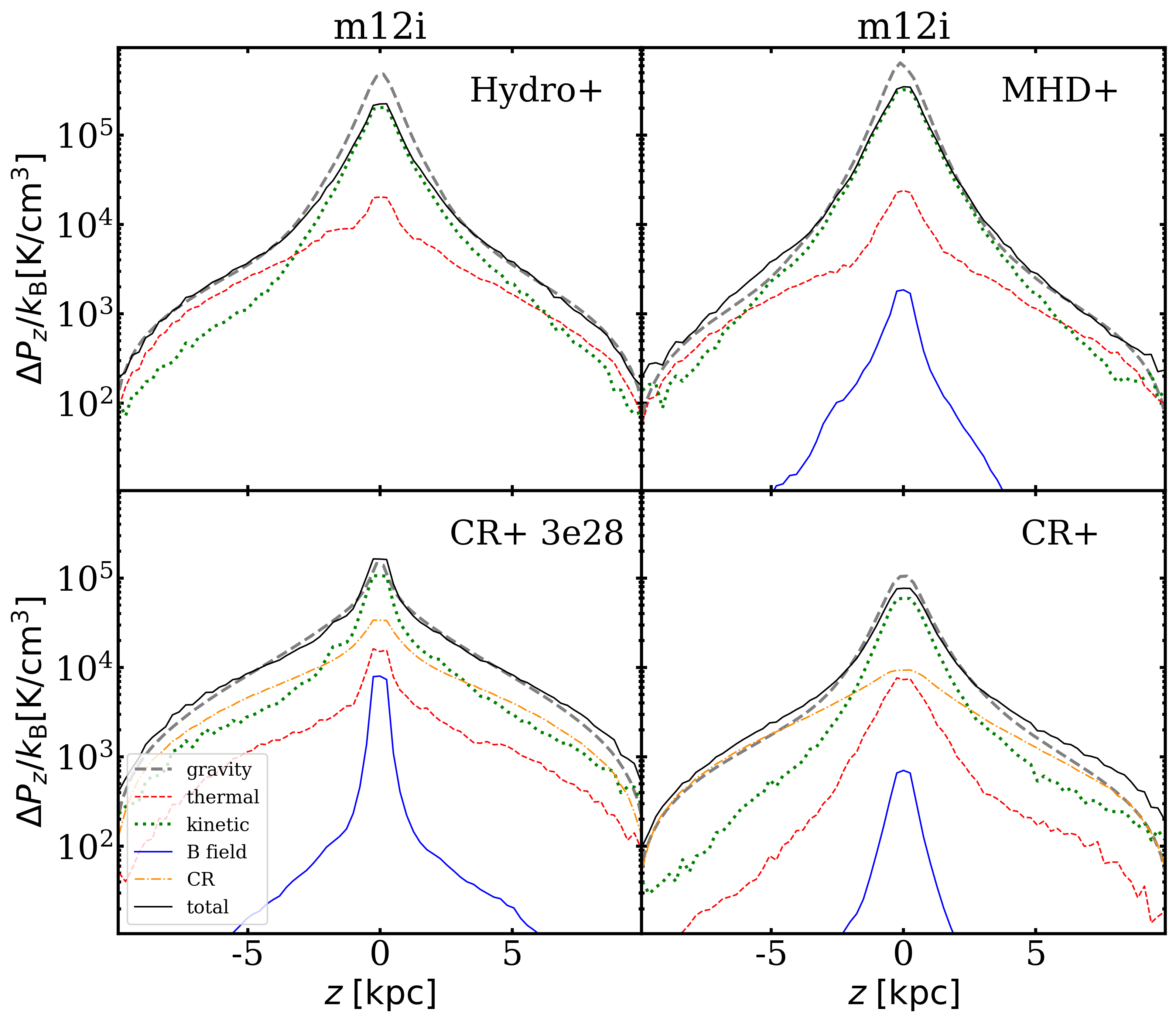}
  \includegraphics[width={0.75\textwidth}]{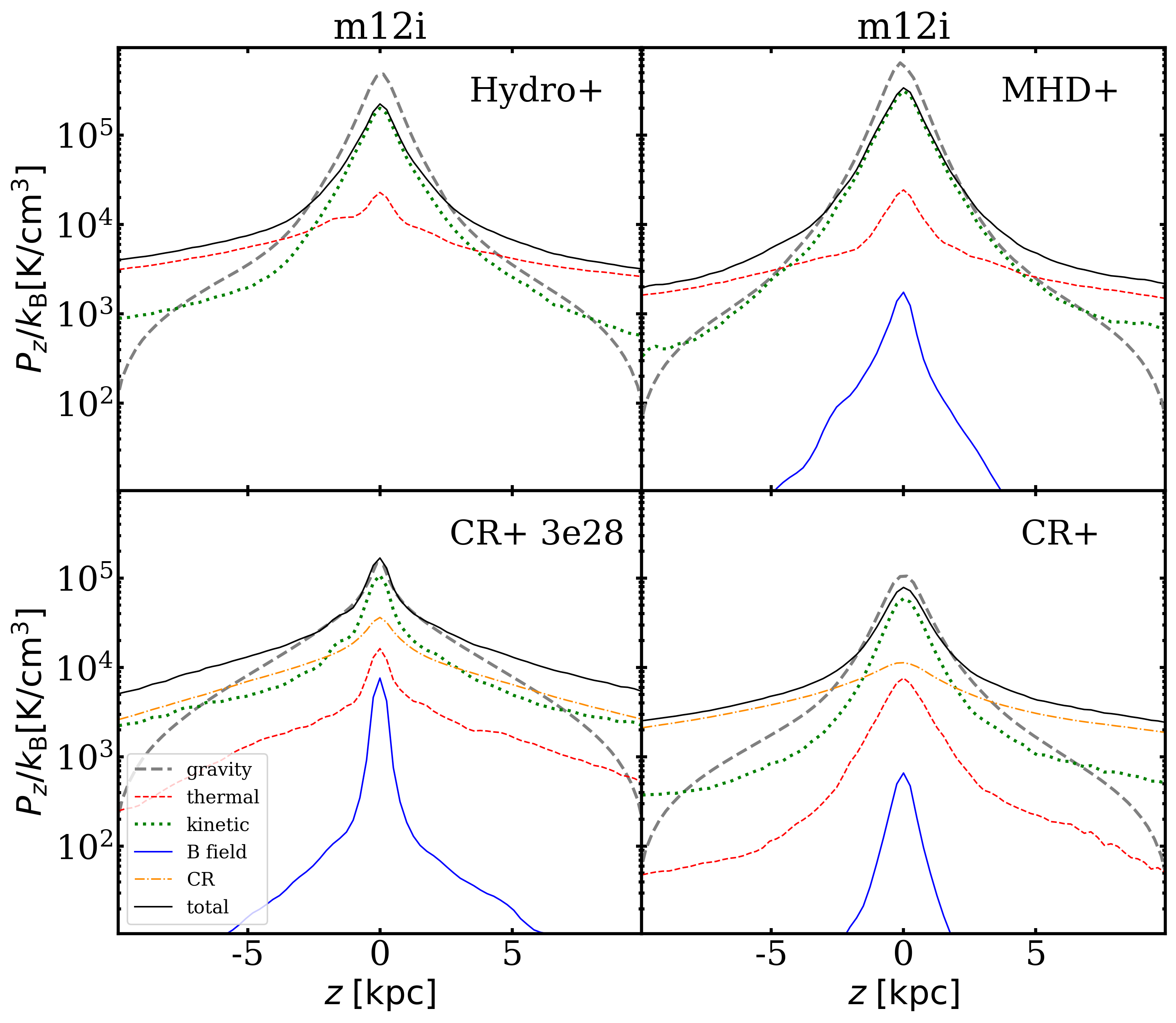}
\caption{({\it Top:})Pressure balance as a function of height (as  Fig.\ref{fig:prebal}). Hydro+ and MHD+ are very similar due to the weak magnetic field support, whereas CR+3e28 has weaker CR pressure support than CR+(3e29). {\it Bottom}: similar plots but we do not subtract the values at z = 10 kpc for pressure, so the pressures here are absolute.}
\label{fig:prebalhydromhdcr}
\end{figure*}

In the main text, we compare the FIRE simulations with hydrodynamics and CR feedback. Here we add comparisons with simulations with magneto-hydrodynamics without CRs (MHD+), and with a lower CR diffusion coefficient (anisotropic diffusion coefficient = $3\times 10^{28}{\rm cm/s}$, CR+3e28).

Fig. \ref{fig:hydromhdcrquiver} shows the velocity and temperature of m12i and m12f in MHD+ and with lower diffusion coefficient. Hydro+ and MHD+ appear very similar to each other, but differ from CR+, which has significantly more warm gas and less hot gas. CR+3e28 has more hot gas than CR+, as expected from the weaker CR pressure outside of the ISM.

In Fig. \ref{fig:gasdenTznc_m12crb70} we show the vertical gas distribution in {\bf m12i} simulations. MHD+ has similar behaviours as Hydro+, e.g. the HIM takes over others at large height. On the other hand, CR+3e28 is similar to CR+, except the hot gas is more extended. Volume filling fraction of the hot gas is also significantly higher in CR+3e28 compared to our default CR+ simulation.

The upper panels of Fig. \ref{fig:prebalhydromhdcr} shows the dynamical support of these simulations. All of the runs satisfy approximate dynamical equilibrium. Hydro+ and MHD+ are both supported by thermal and kinetic pressures, whereas CR+ are supported by CRs at a large height $z$. Magnetic tension is unimportant in both MHD+ and CR+ runs. 

CR+3e28 has more extended and stronger thermal and kinetic pressures compared to the default CR+(3e29) with higher diffusivity. This is because the CR pressure is weaker above the disk, since CRs cannot escape the galaxies before significant hadronic loss, which also leads to higher $\gamma$-ray emission. The CR halos are less extended due to slower diffusion. However, dynamical balance still holds.

The lower panels of Fig. \ref{fig:prebalhydromhdcr} shows the absolute pressure (not subtracting the pressure at 10 kpc). The panels emphasize the thermal (CR) pressures are the most extended and dominant components at $z$ > 5 kpc in Hydro+ (CR+).

\end{document}

%% file: introduction.tex
\section{Introduction}
\label{sec:introduction}
An important question in astrophysics is what supports the gas a few kpc above galactic disks, i.e. at the disk-halo interface. Early models for pressure support above the disk included only $10^4\;{\rm K}$ warm gas, which lead to observational and theoretical challenges. First, a thin gas disk is unstable due to convective and Parker instabilities \citep{Park53instability}. Second, this thermal pressure is inadequate to support the significant weight far above the mid-plane \citep{Boul90}.

One possible solution is to have a {\it dynamical} model. \cite{Shap76galacticfountain} proposed that hot gas is injected from superbubbles \citep{Cast75superbubble} that form a {\it galactic corona}, $10^6{\rm K}$ hot gas, at high-galactic-latitude suggested by \citealt{Spit56corona}. The hot gas then cools down and falls back to the disk, forming {\it galactic fountain}. This infalling gas could constitute some of the observed high velocity clouds \citep{Breg80gfhvc}. To explain the amount of hot gas observed in the interstellar medium (ISM), \cite{Norm89} proposed that the SN-driven superbubbles are elongated inside galactic disks and hot gas is released above, namely the {\it galactic chimney} model.

Another type of models assumes {\it hydrostatic} equilibrium and takes into account non-thermal pressure, e.g. magnetic fields and cosmic rays (CRs) \citep{Chev84CRhalomodel,Boul90,Ferr01,Boet16DIGbalance}. These models are justified by the significant non thermal pressure observed in the solar neighbourhood \citep{Bloe87,Bowy95}, and the large scale heights of magnetic fields and CRs \citep{Bloe93CRheight,Beck15Magneticfield,Gren15CRreview}. CR pressure above the disk can also help explain the observed extent of the warm ionized gas layer, e.g. \cite{Giri16CR,Vand18CRWIM}. Furthermore, \cite{Benj02rotHICRpre} suggested that non-thermal pressure, e.g. CR pressure or magnetic tension, can naturally produce ``lagging halos'': extra-planar gas is observed to be lagging behind the galactic rotation \citep{Swat97HIrot,Frat02xrayngc2403}\footnote{But this lagging could be explained with gas inflows \citep{Frat08inflowdig}.}. 

Finally, the {\it CR-driven wind} model suggests that CRs are not only able to balance the gas weight, but also drive galactic winds \citep{Ipav75}.  \cite{Brei91,Brei93} developed more sophisticated 1D analytic models accounting for more realistic magnetic field and galaxy morphologies (see also \citealt{Recc16crgw,Socr08,Mao18,Croc20CRMW,Quat21CRwind}). Notably, models with CR-driven winds (or hybrid thermal-CR winds) are more successful in explaining the mild radial gradient of Milky Way's (MW) $\gamma$-ray emission \citep{Brei02} and the soft X-ray emission \citep{Ever08} than the hydrostatic models. CR-driven winds have been subsequently explored in many ISM, isolated galactic disk, and cosmological simulations (e.g. \citealt{Hana09,Boot13,Sale14,Giri16CR,Pakm16,Rusz17,Hopk20CRwind,Bust20CRLMC,Bust21crism,Buck19crMW}).  

In reality, gas around galaxies is likely a mixture of these models \citep{Cox05}. For example, many X-ray emitting hot bubbles are observed in late type galaxies \citep{Tyle04xrayspiral}. Many spiral galaxies also emit strongly in radio, suggesting the existence of strong CR and/or magnetic field pressures \citep{Tull00nonthermal}. 

Furthermore, we must also consider the interactions of outflows with circumgalactic medium (CGM) and infalling gas \citep[e.g.][]{Kere05, Mura15, Angl17barcycle, Hafe19CGMorigin}. Infalling gas and the CGM can co-rotate with galactic disks \citep{Kere09c,Stew17coldflow} and dynamically affect outflows and fountains \citep{Frat08inflowdig}. Cosmological simulations provide natural and suitable boundary conditions \citep{deAv04,Hill12,Mart16,Fiel17SNwind} for modeling the disk halo interaction.


Consequently, the support of gas above galactic disks is best studied with cosmological galaxy simulations with magnetic fields and CRs. However, simulating CR propagation (diffusion and streaming) was thought to be extremely computationally expensive at high resolution \citep{Shar10str,Shar11aniso}. Recently, \cite{Jian18}, \cite{Thom19}, and \cite{Chan18cr} developed computationally efficient algorithms for CR diffusion {\it and} streaming. Using the algorithm developed in \cite{Chan18cr}, \cite{Hopk19cr} performed and analysed an extensive suite of high resolution cosmological galaxy simulations with magneto-hydrodynamics (MHD) and CRs. In this work, we select the set of CR runs from \cite{Hopk19cr} calibrated to reproduce the observed $\gamma$-ray emission to gas density relation \citep{Chan18cr,Hopk19cr}, which also matches the MW grammage constraint \citep{Hopk20CRtrans}. They yield a better match to the cool CGM properties \citep{Ji19CRCGM} when compared to simulations without CRs.


In this paper, we investigate the disk-halo interaction of $L\star$ galaxies with the high resolution cosmological simulations with/without CRs \citep{Hopk19cr}. These simulations also include a comprehensive set of stellar feedback processes, including stellar winds, radiation, and supernovae \citep{FIRE2}. In particular, the momentum imparted by SNe to the surrounding gas is calculated in a resolution insensitive way. Collective effects of spatial and temporal clustering of stellar feedback naturally lead to strong galactic outflows that regulate growth of galaxies on cosmological timescales \citep{Mura15,Mura17}. 


The paper is organized as follows. In \S \ref{sec:method}, we start with a brief description of the simulation code, simulated physics, simulation suite, and analysis methods. In \S\ref{sec:vertical}, we compare different pressure components and assess the dynamical equilibrium with and without CRs. We investigate the effects of CR feedback on the gas flows around galaxies in \S\ref{sec:vergasflow}, and the multiphase gas distribution in  \S\ref{sec:verticaldensity}. Finally, we point out the caveats, some implications on the {\it global ISM} model, and the future work in \S\ref{sec:discussion} and \S\ref{sec:conclusion}.

%% file: method.tex
\section{Methods}
\label{sec:method}
\subsection{Simulation code, simulated physics, and simulation setup }
\label{sec:simulations}
The simulations used in this analysis are first introduced in \cite{Hopk19cr}, where the details of simulated physics and numerical schemes can be found. We briefly summarize their key ingredients below. 

\subsubsection{Simulated physics}
\label{sec:simphy}
We simulate with the {\small GIZMO}\footnote{http://www.tapir.caltech.edu/$\sim$phopkins/Site/GIZMO} code \citep{Hopk15} in the Lagrangian mesh-free finite mass (MFM) mode for hydrodynamics. This Lagrangian Godunov-type method has the advantage of both Lagrangian (smoothed particle hydrodynamics) and Eulerian methods. It captures shocks and fluid mixing instability accurately, allows adaptive spatial resolution, and conserves energy and momentum.

{\small GIZMO} uses an updated version of the PM+Tree algorithm from Gadget-3 \citep{Spri05} to calculate gravitational forces and adopts fully conservative adaptive gravitational softening for gas \citep{Pric07}.  We apply the redshift-dependent and spatially uniform ultraviolet (UV) background model from \cite{Fauc09} that ionizes and heats gas in an optically thin approximation and use an approximate prescription to account for self-shielding of dense gas. We include atomic, metal-line, and molecular cooling for $T=10-10^{10}$ K  with the tabulated cooling rates from CLOUDY \citep{Ferl13}. We also include the explicit treatment of turbulent diffusion of metals,  as in \cite{Colb17} and \cite{Esca18}.

Star formation and stellar feedback are implemented with the FIRE-2 scheme \citep{FIRE2}\footnote{\url{http://fire.northwestern.edu/}}, an updated version of the FIRE model \citep{FIRE}. Stars form in dense ($n_{\rm H} \geq 1000\, {\rm cm^{-3}}$), self-gravitating, molecular gas with 100\% instantaneous star formation efficiency (SFE) per local free fall time \footnote{ We emphasize that 100\% SFE applies only to dense, molecular, self-shielding, and locally-self-gravitating gas, while we set 0\% SFE elsewhere. \cite{Hopk13densemoleculargas} and \cite{Orr18} showed that our simulations with this star formation algorithm match the observed effective star formation efficiency in giant molecular clouds (GMCs) (1-10\%) \citep{Bigi08SFEsubkpc,Lero08SFE,Lee16SFEGMC}. Observations (e.g. \citealt{Kauf13virialparamcloud}) show that most of the gas in massive GMC complexes has a local virial parameter $>1$, so only a small fraction of this gas would satisfy our SF criteria.}

After stars form, we follow stellar feedback, including stellar winds, Type II and Type Ia supernovae (SNe) \citep{Hopk18sne}, and radiation (photoelectric and photo-ionization heating, UV/optical/IR radiation pressure) \citep{Hopk20Rad}. We use the Kroupa initial mass function (IMF) \citep{Krou02} and the STARBURST99 stellar population synthesis model \citep{Leit99} to calculate the energy, momentum, mass and metal return from stars.

We model magnetic fields with idealized magnetohydrodynamics (MHD) following \cite{Hopk16MHD}. Fully anisotropic Spitzer-Braginskii conduction and viscosity are implemented as described in \cite{Hopk17diff} and \cite{Su17}. 
The numerical implementation of CR physics is described in \cite{Chan18cr}. We approximate CRs as an ultra-relativistic fluid in a ``single energy bin'' approximation with adiabatic index $\gamma_{\rm cr}=4/3$. We assume strong coupling between CRs and gas, so CRs also contribute to the the effective sound speed.

We inject 10\% of SNe (II \& Ia) kinetic and stellar wind energy into CRs, which can dissipate and heat the surrounding medium through hadronic and Coulomb interactions \citep{Guo08}. CRs can transport via advection, anisotropic diffusion with a constant diffusion coefficient of $3\times 10^{29}{\rm cm^2/s}$ calibrated to match the relation between total $\gamma$-ray luminosity and gas density \citep{Chan18cr}\footnote{The diffusion coefficient of the order $10^{29}{\rm cm^2/s}$ near galactic disk is also favored by \cite{Evol18CRGH}. They derived similar order of magnitude diffusion coefficients (for GeV CR protons) from wave self-generation and advection as the best fit for the CR spectra.}. We also consider Alfvenic streaming with streaming speed equal to the Alfven speed ($v_{\rm st}= v_{\rm A}$) and streaming losses equal to $\Gamma_{\rm st}=-{\bf v}_{\rm A}\cdot\nabla P_{\rm cr}$. CR transport is implemented with a two-moment method \citep{Chan18cr}, which is computationally efficient and can simultaneously simulate diffusion and streaming (see also \citealt{Jian18,Thom19}).

We consider a standard flat $\Lambda$ cold dark matter cosmology with $\Omega_0\sim 0.27$, $\Lambda\sim0.73$, $b\sim0.045$, and $h\sim0.7$. The initial conditions (ICs) of the runs are generated with the {\small MUSIC} code \citep{Hahn11}\footnote{The configuration files for the IC generation are publicly available here \href{http://www.tapir.caltech.edu/~phopkins/publicICs/}{http://www.tapir.caltech.edu/$\sim$phopkins/publicICs/}}. To achieve high resolution to resolve the multiphase ISM and incorporate our feedback model, we adopt the ``zoom-in'' technique\citep{Port85,Katz93,Onor14}.

\subsubsection{Simulation Suite}
In this paper, we consider three variations of physics and adapt notations used in \citealt{Hopk19cr}: 
\begin{itemize}
    \item Hydro+: default FIRE-2 physics \citep{FIRE2} without magnetic fields or CRs;
    \item MHD+: with all of the physics described in \S \ref{sec:simphy}, except CRs;
    \item CR+: with all of the physics described in \S \ref{sec:simphy}, including CRs;
\end{itemize}

We select three simulated galaxies close to the MW-mass (simulated $M_*\sim 3-9 \times 10^{10}\msun$) from \cite{Hopk19cr}. Their properties are summarized in Table \ref{tab:SIC}. The initial conditions of {\bf m12i}, {\bf m12f}, and {\bf m12b} are identical to those in the  \cite{Wetz16,FIRE2} simulations. Note that the ``Hydro+'' runs in general have higher stellar masses and star formation rates (SFRs) than the ``CR+'' runs. {\bf m12f} has slightly higher SFRs around $z=0$ caused by the interaction and gas accretion from a nearby satellite.

We emphasise that $\langle \epsilon_{\rm gas} \rangle^{\rm sf}$ is our adopted minimum allowed value of the gravitational force softening length under the adaptive softening scheme. In practice, this softening length is not reached, e.g. the minimum gas smoothing length and corresponding softening are around 8 pc in the m12i CR+ run. Therefore, we cannot resolve the {\it thermal} Jeans scale of gas around or above the star formation threshold of 1000 ${\rm cm^{-3}}$\footnote{ Since we don't resolve the thermal Jeans length and related (physical) fragmentation for individual stars, we are using a model for the IMF in our stellar particles \citep{FIRE2}. }. 

But even if the simulations do not resolve the thermal Jeans scale of cold dense gas, our Lagrangian method does not have artificial fragmentation of differentially rotating disks in the sense of \cite{True97artfrag} (unlike in grid codes), as long as the consistency between gravitational and hydrodynamic resolution is maintained \citep{Bate97artfragSPH,FIRE2,Grud21starforge,Yama21artfrag}. However, physical fragmentation might be prevented or slowed down if the thermal Jeans scale is not resolved, i.e. ``under-fragmentation’’.

Nonetheless, the thermal Jeans scale is only a conservative resolution requirement for fragmentation, since thermal pressure is sub-dominant compared to kinetic (or turbulent) and CR pressures (see below). In particular, \cite{FIRE2} showed, even though the thermal Jeans mass in cold gas is not resolved, the GMC mass function is convergent at the resolution we adopt.

\begin{table*}
\centering
    \begin{tabular}{llllllllll}
    \hline\hline
    Name & $M_{\rm vir}$&$M_{*}^{\rm Hydro+}$& $M_{*}^{\rm CR+}$&${\rm SFR}^{\rm Hydro+}$&${\rm SFR}^{\rm CR+}$  & $m_{i,\,1000}$ & $\langle \epsilon_{\rm gas} \rangle^{\rm sf}$        \\
    & $[10^{12}\msun]$&  $[10^{10}\msun]$  &   $[10^{10}\msun]$&  $[{\rm\msun/yr}]$ &  $[{\rm\msun/yr}]$ &$[10^3\,\msun]$ & $[{\rm pc}]$ \\
    \hline
    	\hline
{\bf m12i} & 1.2&7.4  & 2.6&7.8 &2.3& 7.0 & 2.0\\
{\bf m12f} & 1.6&8.9 & 3.6&10.2 &4.7& 7.0 & 1.9\\
{\bf m12b} & 1.3 & 9.1 &3.4&8.5  &2.3&7.0  & 2.2\\
    \hline
    \hline
    \end{tabular}
 	\caption{Simulation details. $M_{\rm vir}$ is the virial mass (with the definition from \citealt{Brya98}); $M_{*}^{\rm Hydro+}$ and $M_{*}^{\rm CR+}$ are the stellar masses in the {\rm Hydro} and {\rm CR+} runs (within 15 kpc from the halo center); SFR is the star formation rate averaged over 100 Myr; $m_{i,\,1000}$ is the mass resolution of gas; and $\langle \epsilon_{\rm gas} \rangle^{\rm sf}$ is the minimum gravitational force softening length.} 
\label{tab:SIC}
\end{table*}

\subsubsection{Halo finding and galaxy orientation}
\label{sec:AHFhalo}
We locate halos and calculate their virial masses and radii with the Amiga Halo Finder ({\small AHF}) \citep{Knol09}\footnote{http://popia.ft.uam.es/AHF/Download.html}. AHF locates the prospective halo center with an adaptive mesh refinement hierarchy. We use the {\small MergerTree} code in {\small AHF} to follow the main progenitor of a halo and study its time evolution (whenever mentioned in the text). We determine the virial over-density and radius with the \cite{Brya98} formulae and define the halo mass as the total mass within the virial radius. 

We rotate the galaxy such that the galactic midplane is perpendicular to the total angular momentum of stars within 5 kpc from the halo center. To locate the galactic center, we place stars (within 5 kpc from the halo center) onto $50\times50\times50$ resolution mesh and define the highest density peak as the galactic center. 

\subsection{Theoretical framework for dynamical balance}
\subsubsection{Volume-weighted quantities}
To calculate the volume-weighted quantities $\left \langle X \right \rangle$, we divide the simulation volume around a galaxy into cubes with side length $l_{\rm cube}$ (=0.25 kpc in the fiducial case). We consider the volume weighted quantity inside a cube $j$ (with a volume $V_j$) to be:
\begin{align}
\left \langle X \right \rangle_j= \frac{\int X{\rm d}V}{\int {\rm d}V}=\frac{\sum _i X_im_i/\rho_i}{\sum _i m_i/\rho_i},
\label{eq:Pvolw}
\end{align}
where $X_i$ is the quantity X of particle i and we sum over all particles in $V_j$, so $\left \langle  \right \rangle_j$ denotes volume-average within cube j. 

\subsubsection{Dynamical Equilibrium: Definition}
\label{sec:hydrobal}

We define ``dynamical balance'' following e.g. \citet{Boul90}. In an appropriate inertial Cartesian frame (which we take to be the center-of-momentum frame of the galaxy, with the $\hat{z}$ axis defined to point along the total angular momentum vector of the galaxy), take the kinetic MHD momentum equation in conservative form (assuming tight-coupling of CRs), and assume local steady-state (so the Eulerian time derivatives vanish), giving
\begin{align}
\label{eqn:momentum}\frac{\partial \rho\,{\bf v}}{\partial t} &=
-\nabla \cdot \left[ \rho\,{\bf v}\otimes{\bf v} + \mathbb{P}_{\rm thermal} + \mathbb{P}_{\rm cr} +  \mathbb{P}_{\rm B} \right] + \rho{\bf g}
\rightarrow 0
\end{align}
where $\rho$ is the gas density, ${\bf v}$ gas velocity, $\mathbb{P}_{\rm thermal}$ the thermal pressure tensor which we take in our analysis be $P_{\rm thermal}\,\mathbb{I}$ with $P_{\rm thermal} = n\,k_{B}\,T$ (we ignore the anisotropic Braginskii kinetic terms here, as they are generally small), $\mathbb{P}_{\rm cr}$ the CR pressure tensor (which we again approximate as isotropic, $=P_{\rm cr}\,\mathbb{I}$, with $P_{\rm cr}=e_{\rm cr}/3$), $\mathbb{P}_{\rm B} \equiv (B^{2}\,\mathbb{I}-2\,{\bf B}\otimes{\bf B})/8\pi$ is the magnetic pressure tensor, and ${\bf g}$ is the gravitational acceleration. 

Then we take the $\hat{z}$ component of Eq.~\ref{eqn:momentum}, and average it over thin ``slab'' between $z\rightarrow z+dz$, assuming that mass flux in the $xy$ directions and the cross terms (e.g. off-diagonal terms) between $v_{xy}$ and $v_{z}$ (or $B_{xy}$ and $B_{z}$) are negligible after integration, giving:
\begin{align}
\frac{\partial }{\partial z}\left[
\langle P_{\rm thermal} \rangle + \langle P_{\rm cr} \rangle + \langle \Pi_{\rm B} \rangle + \langle P_{\rm kin} \rangle \right] = -\langle \rho g_{z} \rangle
\label{eq:pressurebalance}
\end{align}
where $\Pi_{B} \equiv (B^{2}-2\,B_{z}^{2})/8\pi$ is the magnetic tension and $P_{\rm kin} \equiv \rho\,v_{z}^{2}$ is the kinetic pressure, and $\langle X \rangle$ refers to the volume-average of quantity $X$ within the {\em slab}. Note that we have omitted the momentum source terms because the momentum source terms are shown to be small away from galactic midplane and galactic center \citep{Gurv20prebal}. 

In order to calculate the gravity and other pressure components, we further divide the slabs into cube $j$. Note that the {\em kinetic} component
$\langle P_{\rm kin}\rangle_j$ can be separated into a local {\em bulk flow} component $\langle P_{\rm bulk}\rangle_j \equiv \langle\rho\,\bar{v}_{z}^{2}\rangle_j$ (due to the mass weighted mean velocity within the cube j, $\bar{v}_{z}$) and a {\em dispersion} component $\langle P_{\rm disp}\rangle_j \equiv \langle\rho\,\sigma_{z}^{2}\rangle_j$ with $\sigma_{z}\equiv v_{z}-\bar{v}_{z}$, i.e.
\begin{align}
\langle P_{\rm kin}\rangle_j=\langle P_{\rm bulk}\rangle_j+\langle P_{\rm disp}\rangle_j.
\end{align}
We stress that the {\em dispersion} term simply includes {\em all} kinetic energy of non-uniform motion -- we make no effort to actually determine if this motion is {\em turbulent} in any rigorous sense, and it should not be interpreted as such.

The right hand side shows the gravity, where $g_z=|{\bf g}\cdot\hat{\bf z}|$ is the magnitude of gravity in the vertical direction. The sum of the left hand side is the total support ({\em total}), which will be equal to the gravity in dynamical balance.

Our approach is nearly equivalent to \cite{Gurv20prebal}, except that we include all of the gas in the calculation, whereas they exclude dense gas. Since we are focusing more on pressure above the disk, this difference should not be important. Furthermore, \cite{Gurv20prebal} deposited gas particles into meshes according to their kernels, but we deposit according to their locations. The difference is insignificant if we average over galactic disks and tens of Myrs, as we do in the next section.

This definition of dispersion pressure is also similar to other simulation work, e.g. \cite{Joun09ISMtur} (although they took the mean of velocity dispersion in three orthogonal directions). On the other hand, \cite{Kim18solarneigh} adapted $\rho v^2_\mathrm{z}$ in the {\em turbulent pressure}, so their {\em turbulent pressure} is what we refer to as the total {\em kinetic pressure}.

Note that the dispersion pressure is a function of the cube size $l_{\rm cube}$, e.g. it is weaker with a smaller grid \citep{Gurv20prebal}. We pick the fiducial cube length $l_{\rm cube}$ to be 0.25 kpc because it is close to the turbulence generation scale \citep{Elme03,Bour10tur} and \cite{Joun06} found most of the turbulence energy is within 200 pc scale. But we also show the dispersion pressure for other cube sizes in the next section.

We define {\it hydrostatic} balance by pressure balance excluding the bulk flow pressure, and {\it dynamical} balance by including the bulk flow pressure term (consistent with the terminology in \citealt{Gurv20prebal}).

Some of the non-hydrostatic solutions, e.g. with steady inflows/outflows/fountains, can be still in dynamical balance, where the bulk flow pressure helps to {\it balances} gravity, i.e. ${\rm d}\left(\left \langle P_{\rm bulk}\right \rangle+\left \langle P_{\rm other}\right \rangle\right)/{\rm d}z \approx -\left \langle\rho g_{\rm z}\right \rangle$. In our definition, dynamical equilibrium means that the system is in a steady state, at least in the vertical direction. Note that this terminology is different from some literature \citep[e.g.][]{Hill12,Boet16DIGbalance,Boet19DIGbalance}.

We estimate the gravitational acceleration $\langle{\bf g}\rangle_j$ at the center of the cube $j$ by summing over only particles within 100 kpc from the galactic center and pick one particle from every hundred, but increase its mass by one hundred times\footnote{Comparing the gravitational force without this approximation, we found that the error is $\sim$ 10\% around the midplane and even smaller above the plane.}.

\subsection{Multiphase medium and temperature cut}
\label{sec:multiphasedef}
We divide gas into five phases, the cold medium (CM; $T<5\times10^3{\rm K}$), warm neutral medium (WNM; $5\times10^3{\rm K}<T<2\times10^4{\rm K}$), warm ionized medium (WIM; $5\times10^3{\rm K}<T<2\times10^4{\rm K}$), warm hot ionized medium (WHM; $2\times10^4{\rm K}<T<5\times10^5{\rm K}$), and hot ionized medium (HIM; $T>5\times10^5{\rm K}$). Note that CM includes both molecular gas and cold neutral gas. The WHM is the transitional phase, which means gas is unlikely to stay in this phase for a long time under pure thermal pressure. Note that these temperature cuts are different from \cite{Gurv20prebal}, but similar to \cite{Vija20verbal}, who performed a similar analysis on the {\small TIGRESS} simulations.

\section{Conceptual Models of the disk-halo interaction}

\begin{figure}
\includegraphics[width={0.46\textwidth}]{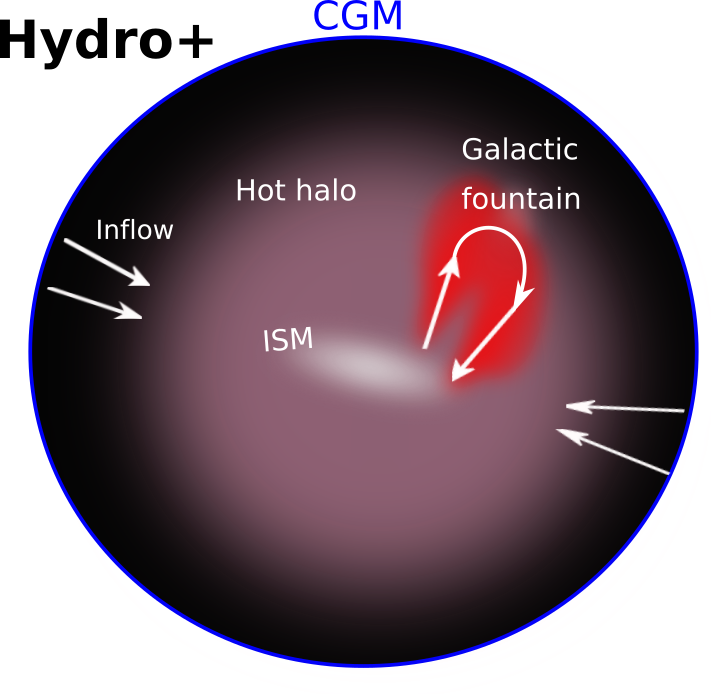}
\includegraphics[width={0.45\textwidth}]{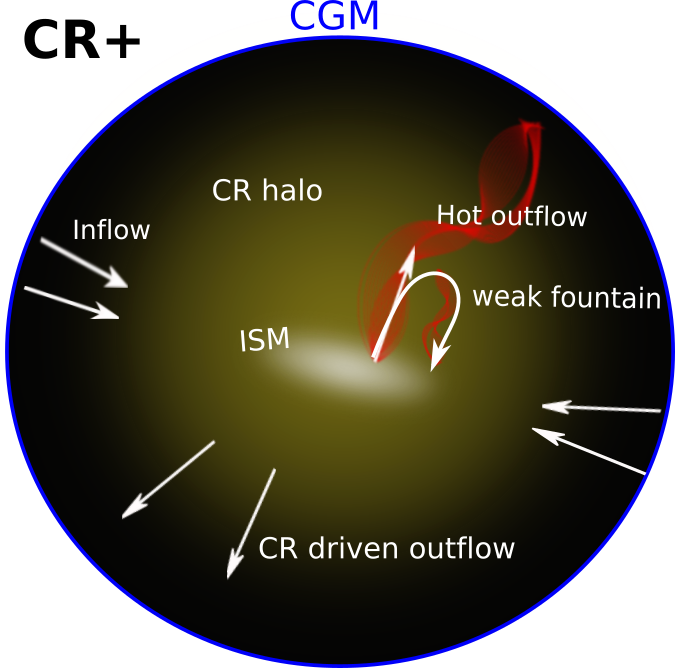}
\caption{Schematic illustrations of the inner gaseous halos in the Hydro+ and CR+ runs. A recently formed super-bubble drives hot gas, shown as red regions. White areas represent the ISM, dominated by kinetic pressure. Red areas show the thermal-pressure dominated regimes, whereas yellow areas represent CR-pressure dominated regimes.}
\label{fig:schematic}
\end{figure}

In the introduction, we mentioned different models for the disk-halo interface. Based on our results in Section \ref{sec:results} we construct the following conceptual model of the zone between galactic ISM and the CGM. The models differ significantly for Hydro+ and CR+ and is illustrated in Fig. \ref{fig:schematic}. 

Hydro+ is best described with a ``galactic fountain'' model, where superbubbles break out above the disk and, once cooled, the gas falls back to the disk. Hot gas from the bubbles is not able to escape from the halo \citep{Mura15}, possibly due to strong gravity and hot gas confinement (see discussion in \S\ref{sec:outflowfate} and \citealt{Hopk20CRwind}). Additional infall of hot gas is supplied from the hot CGM (produced by virial shocks; see \citealt{Ji20cr}). 

The CR+ simulations can be described with a hybrid model with CR-driven wind and hot gas bubbles. Around the galactic disks, there are hot super-bubbles and gas is supported by kinetic pressure. At a larger height, gas is supported or even pushed out by CR pressure, so the galaxies are embedded in a WIM/WHM halo. CR pressure dominates in the halo and prevents rapid recycling of fast winds (i.e. there are only weak galactic fountains), but instead, allows fast SN heated gas to escape from the inner halo.

While not explored in this work, the bulk of the infalling gas in our simulated galaxies accretes radially in the disk midplane and is not contributing significantly to the vertical gas flows (see \citealt{Hafe19CGMorigin,Trap21gasinflow}.

We have presented two distinctive conceptual models for no CRs and CR-dominated cases, but reality might lie between these two cases.
For example, \cite{Ji19CRCGM} found that the no-CR simulations were in tension with observed OVI absorption, while the CR-dominated model had difficulty reproducing observations for high ions.

%% file: results.tex
\section{Results}
\label{sec:results}
\subsection{Vertical Support and Dynamical Equilibrium}
\label{sec:vertical}

\begin{figure*}
 \includegraphics[width={0.95\textwidth}]{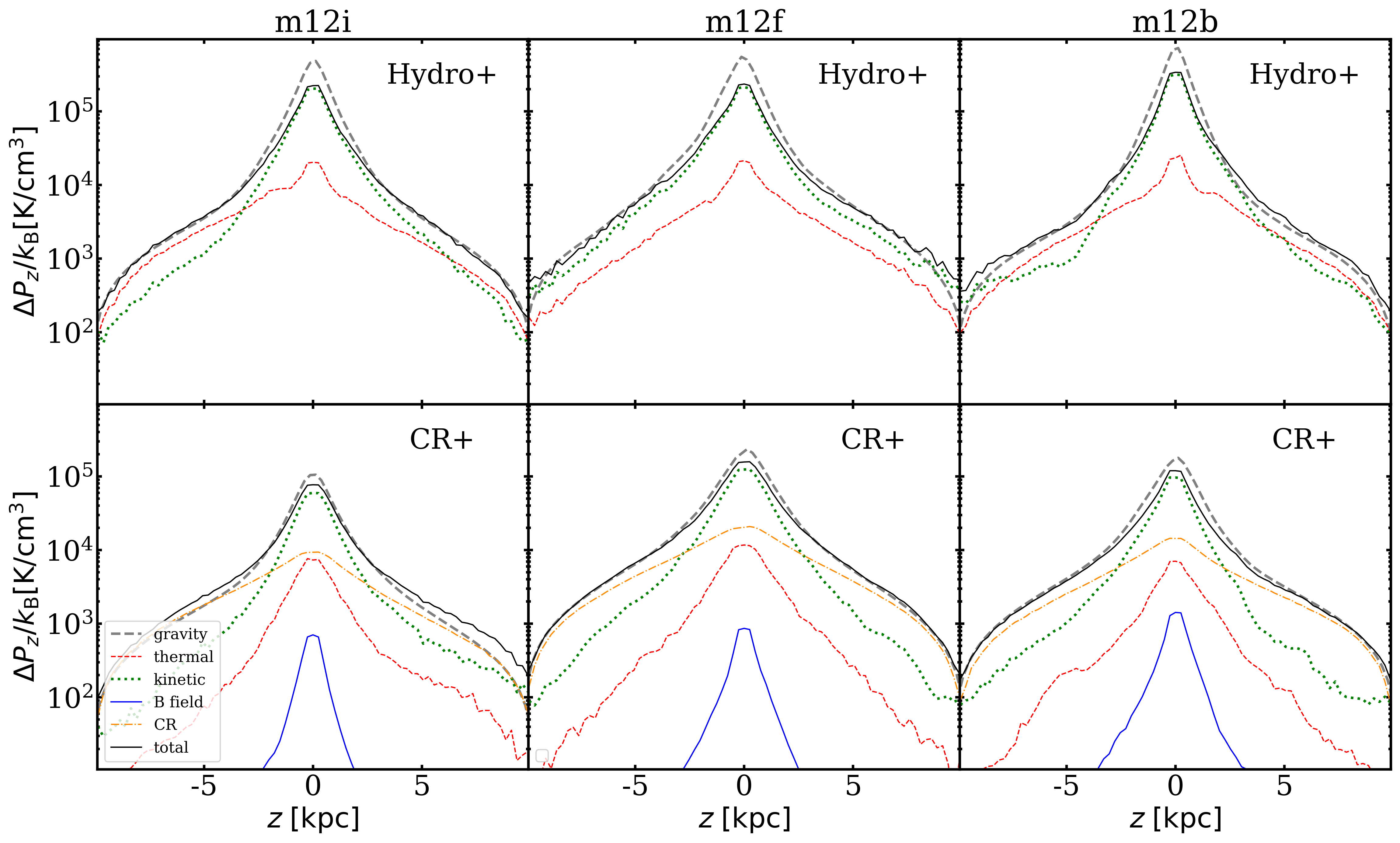}
\caption{Vertical support from different components averaged within $R = 9\; {\rm kpc}$ and over the last 250 Myr, as a function of height above the disk, $z$. $\Delta P_z$ is the vertical pressure difference between $z$ and $z =10\; {\rm kpc}$ (note that this difference can quickly approach zero at heights close to z =10 kpc, since we subtract the $z =10\; {\rm kpc}$ value). {\em total} represents the sum of thermal, kinetic, CR, and B field supports (see Eq. \ref{eq:pressurebalance} for their definitions). Note that $7\times 10^3\;k_{\rm B}\;{\rm K/cm^3}\approx10^{-12}\,{\rm dyne/cm^{2}}$. The slopes of gravity steepen around 2 kpc because the baryonic disk dominates the gravity below 2 kpc. Total pressure roughly balances the gravity $>1 \;{\rm kpc}$ from the midplane, so dynamical balance is satisfied.}
\label{fig:prebal}
\end{figure*}

\begin{figure}
 \includegraphics[width={0.47\textwidth}]{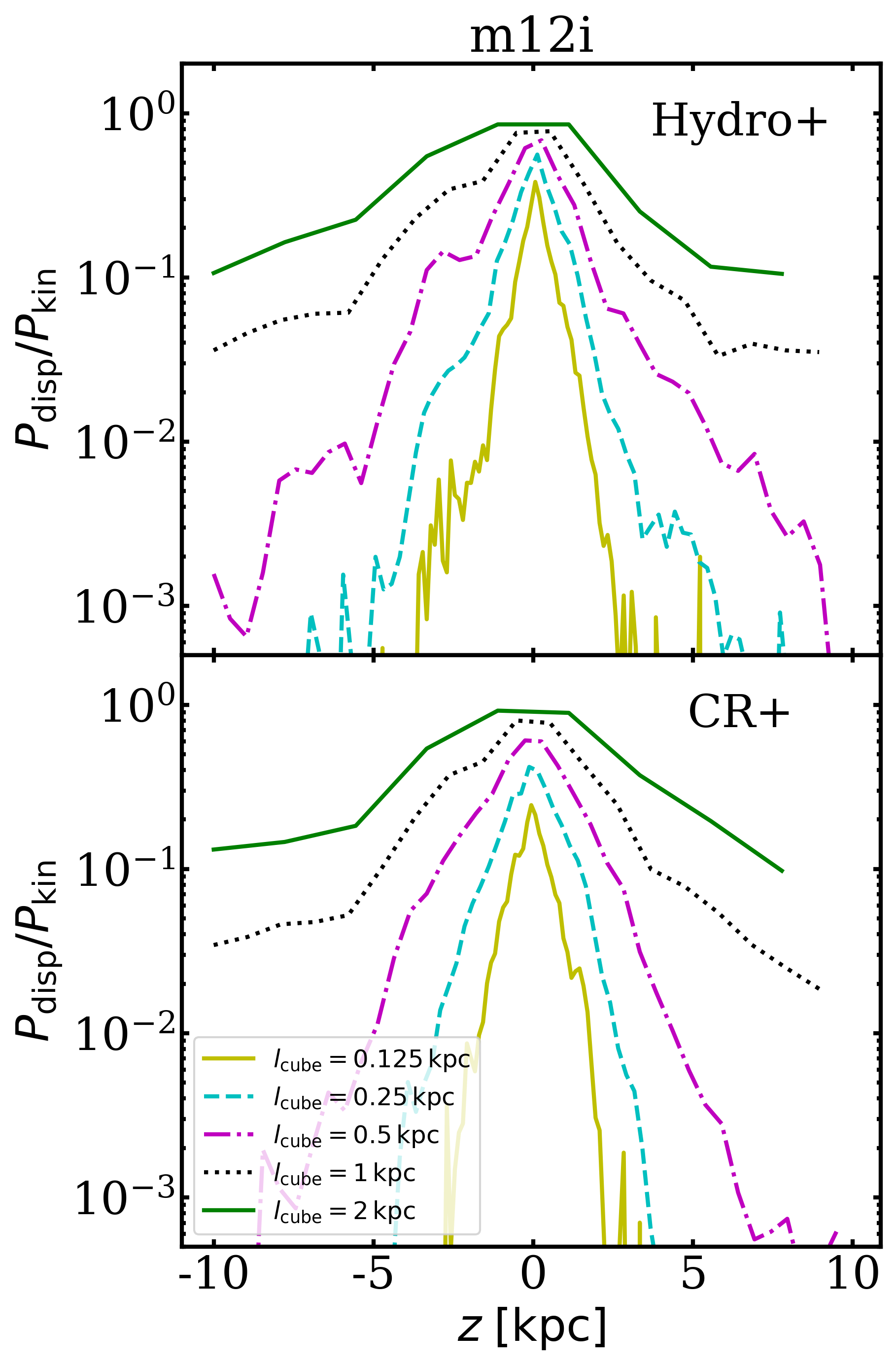}
\caption{The fraction of total kinetic pressure $\langle P_{\rm kin}\rangle$ in dispersion component $\langle P_{\rm disp}\rangle$ averaged over different local box size, $l_{\rm cube}$, in {\bf m12i}. Varying the box size shows that most of the kinetic pressure at the midplane is on scale of $\sim$kpc.}
\label{fig:pztur_ke}
\end{figure}

Fig. \ref{fig:prebal} compares different pressure components from Eq. \ref{eq:pressurebalance} averaged over the last 250 Myr. Specifically, we integrate ${\rm d}\langle P\rangle/{\rm d}z$ from $\pm10\;{\rm kpc}$ to a given height $z$ to get the pressure (tension) difference $\Delta P$, or equivalently take the pressure difference between the gas at the height $z$ and $10\;{\rm kpc}$ away from the mid-plane. Note that this subtraction can be lead to very low values of pressure difference. For example, Fig. \ref{fig:prebalhydromhdcr} shows that difference in thermal or CR pressure can drop by a factor of a few near $z =10\;{\rm kpc}$ after the subtraction, since thermal or CR pressure vertical profile can be very flat.

\begin{figure*}
 \includegraphics[width={0.95\textwidth}]{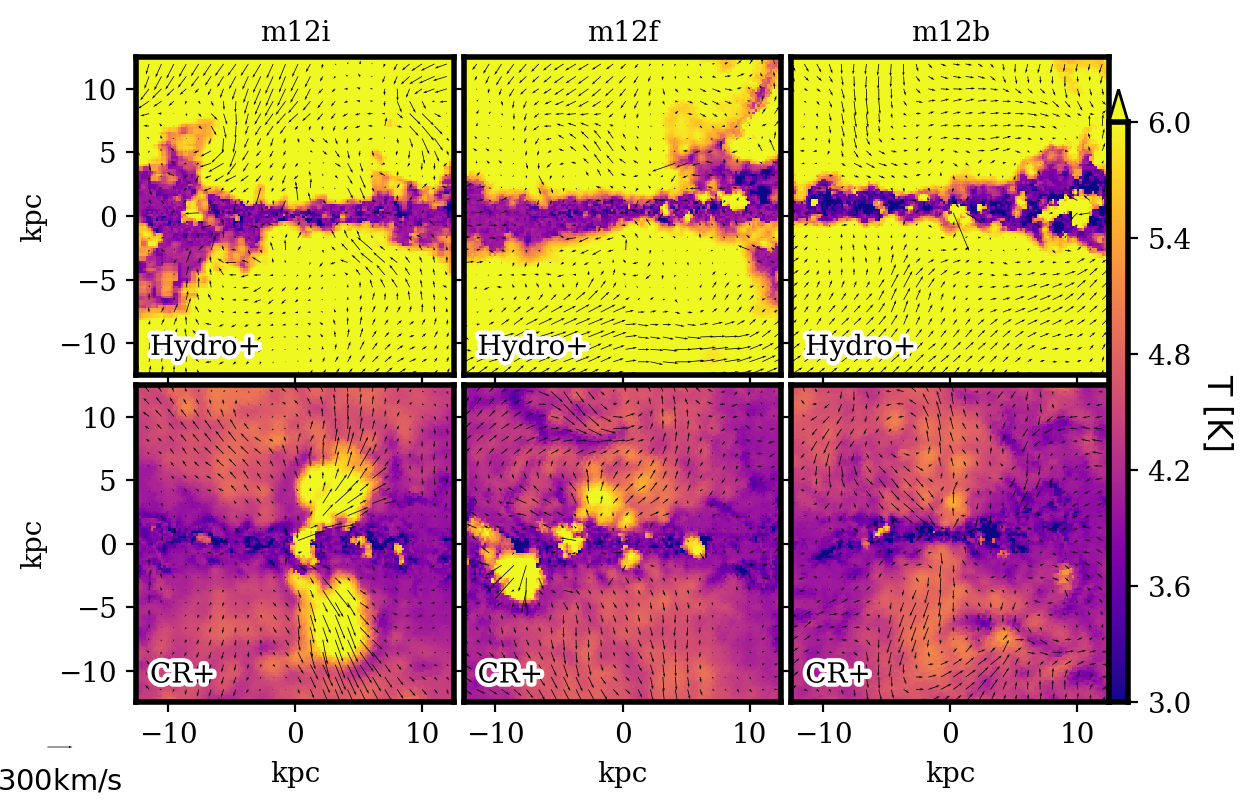}
\caption{Edge-on slice plots of mass-weighted temperature profiles of the {\bf m12}s galaxies at redshift $=0$, where arrows represent gas velocities. The gas temperature above galactic disk is significantly lower with CRs, especially away from the midplane. The CR runs show slow uniform outflows driven by CR pressure gradient, in addition to intermittent, fast hot SN driven outflows. 
}
\label{fig:slicem12tempvel}
\end{figure*}

In both Hydro+ and CR+ runs, we find a good balance from $z =1\;{\rm kpc}$ to $10\;{\rm kpc}$ between vertical support and gas weight (``{\it gravity}''), which implies that the dynamical equilibrium is a reasonable assumption, at least when averaged over 250 Myr. Only in {\bf m12i} CR+, the total pressure is slightly higher than the gravity at heights of several kpc owing to the recent strong superbubbles (see Fig. \ref{fig:slicem12tempvel}). But the deviation is less than a factor of two even in this worst case.

However, dynamical balance appears to be violated around the midplane in all runs likely as a combination of approximations made in our equations and our calculation of physical quantities. 
In deriving the dynamical equilibrium (Eq. \ref{eq:pressurebalance}), following \cite{Boul90} we neglect the large scale correlations between $v_x, v_y$ and $v_z$, e.g. rotation, shear flows, fountains etc, which can be important near the midplane\footnote{Numerically, the gravitational acceleration is averaged over finite-size grids, so the location at which we estimate gravity is slightly mismatched with the gas center of gravity near the midplane; we find the balance is better at smaller grid sizes, but it already converges at the grid size we used, so we use this grid size to reduce computational expense.}.

\cite{Gurv20prebal} also noted an apparent pressure deficit of up to $\sim$ 30\% near the midplane, which can potentially be explained by a combination of effects neglected in the dynamical balance: (a) in general, the left-hand side of Eq. \ref{eq:pressurebalance} should also include an average over momentum source terms, which is non-zero when (as in the FIRE simulations) mechanical and radiative feedback is injected over a finite-size regions; and (b) in detail, our vertical balance analysis neglects correlations between density and velocity components, including motions in the disk plane.

Our midplane pressures are seemingly larger than the MW fiducial value. It is because the usually-quoted MW value ($\sim 10^4 \;{\rm K/cm^3}$; \citealt{Ferr01}) is measured at the solar circle (i.e. $r\sim 9\;{\rm kpc}$). In contrast, Fig. \ref{fig:prebal} includes the central region (from $r=0\; {\rm kpc}$ to $r=9\; {\rm kpc}$), where midplane pressure is much higher. We have explicitly checked that the {\bf m12} midplane pressures are an order of magnitude lower at $r\sim 9\;{\rm kpc}$, so they are not in conflict with the MW value.

Fig. \ref{fig:prebal} and Fig. \ref{fig:prebalhydromhdcr} show B-field does not affect the pressure balance in runs with MHD (with or without CRs). It is because the vertical magnetic tension is very weak, compared to other components, in all of the runs. Note that magnetic tension is only a fraction of the total magnetic pressure (Eq. \ref{eq:pressurebalance}), so it will be weaker than $B^2/(8\pi)$, especially when B field is perpendicular to the disks. \cite{Hill12}, for example, found a good cancellation of $B^2$ and $2B^2_{\rm z}$ terms in their elongated box ISM simulations, so the vertical magnetic tension is very small. \cite{Kim18solarneigh} also found that magnetic field provides insignificant vertical support in their solar neighborhood {\small TIGRESS} simulations. Observations, e.g. \cite{Stei20spiralB}, also suggest B fields have significant perpendicular components (hence a weak tension).

In both Hydro+ and CR+ runs, kinetic pressure is the most dominant component around mid-plane, much stronger than thermal, magnetic, and CR pressures. Kinetic pressure is weaker in CR+ than in Hydro+, since the SFR and the inflow rate are lower (\citealt{Hopk20CRwind,Trap21gasinflow}). 

In Fig. \ref{fig:pztur_ke}, we show that most of the kinetic pressure around mid-plane is from non-uniform gas motion within scale around 1-2 kpc, consistent with what \cite{Gurv20prebal} found. Other pressures play only subdominant roles at the mid-plane. At height > 1 kpc, most of the kinetic pressure instead comes from bulk motions, which we will explore in the next section.

The more prominent difference between Hydro+ and CR+ is the pressure support at a large height. In Hydro+, thermal pressure takes over from kinetic pressure at a height $z\gtrsim 10\;{\rm kpc}$ as expected from hot halos in $L_\star$ galaxies without CRs  \citep{Ji19CRCGM}. However, in CR+, the thermal pressure at a large height is weakened significantly  \citep{Sale16cgm,Buts18,Hopk20CRtrans}). This is because CR feedback modifies the multiphase gas flows and the phase structure of the CGM \citep{Hopk20CRwind,Ji19CRCGM,Ji20cr}. We discuss this in the next section and later show that the near-disk hot gas content is still proportional to the SFR in \S\ref{sec:HIMSFR}, consistent with a star formation driven origin.

Instead of thermal pressure, in CR+, CR pressure takes over at $z\gtrsim 5\;{\rm kpc}$. Thermal and kinetic pressures are strongly suppressed, compared to the Hydro+ runs. Several lines of argument can explain why CR pressure, instead of thermal pressure, can dominate.

First, CRs have a smaller ``effective'' adiabatic index ($\gamma_{\rm cr}=4/3$) than the thermal adiabatic index ($\gamma=5/3$), so CR adiabatic losses are weaker. Second, CRs cool only weakly through hadronic and streaming losses (at low density), and do not suffer from radiative cooling. At the same time, dominant CR pressure slows down gas infall into the halo, lowering post shock temperature and increasing overall density, resulting in increase of radiative losses of the volume filling gas phase and prevention of formation of the hot halo in our $L\star$ galaxies with CRs \citep{Ji20cr}.

A simple estimate can demonstrate that thermal or kinetic pressures are not needed at large $z$ to support the gas, if CRs are present. For our assumption, the CR injection rate is 
\begin{align}
\dot{E}_{\rm cr} = 0.1E_{\rm SN}\frac{\rm SFR}{\langle m_\star\rangle}\xi \sim 3\times 10^{40}{\rm erg/s},
\end{align}
where $E_{\rm SN}$ is the energy per SNe, SFR is star formation rate ($\sim 1 \;\msun/{\rm yr}$), $\langle m_\star\rangle(\sim 0.4\msun)$ is the mean stellar mass, and $\xi=0.0037$ is the fraction of stars that can produce SNe \citep{Krou02}. We assume steady state so that diffusion out of the region is approximately compensated by injection and the losses are small (for high diffusion coefficient)and estimate the CR energy density at 10 kpc \citep{Hopk19cr} as:
\begin{align}
e_{\rm cr} \sim \frac{\dot{E}_{\rm cr}}{4\pi \kappa_{\rm eff} r}\sim 10^{-13}{\rm erg/cm^{3}},
\end{align}
where $\kappa_{\rm eff}$ is the effective isotropically-averaged CR diffusion coefficient ($\sim 10^{29}{\rm cm^2/s}$). 

This CR energy density is comparable to the thermal energy density of halo gas at 10 kpc in our non-CR simulations with $n\sim 10^{-3}{\rm cm^{-3}}$ and $T\sim 10^6{\rm K}$,
so thermal (or kinetic) pressure is not required to support the gas away from galactic disks. In fact, CR pressure can be important even out to virial radii if the diffusion coefficient is constant (see \citealt{Ji19CRCGM,Hopk20CRwind}), effectively supporting a large fraction of the halo gas.

 In the CR-dominated regime, we find that gas density is carefully adjusted to the CR pressure so that the gas weight balances the CR pressure, since hot and rarefied gas will flow out along CR pressure gradient and dense gas will fall in, until the CR pressure gradient balances gravity \citep{Hopk19cr}. 

However, near galactic disks, the gravitational force is strong owing to concentration of baryons. For example, the typical $L\star$ galaxy stellar surface density is 
$\Sigma_\star \sim 200\;\msun/{\rm pc}^2$ and gas surface density is $\Sigma_{\rm gas} \sim 20\msun/{\rm pc}^2$ in our CR+ and Hydro+ simulations \citep{Gurv20prebal,Hopk19cr}. These give the gravitational pressure $\sim G\Sigma_\star\Sigma_{\rm gas}\sim 10^5 {\rm K/cm^3}$, which is larger than the midplane CR pressure $\sim 10^4 {\rm K/cm^3}$, so the additional support from kinetic pressure is required to prevent gas collapse.

Our simulated CR distribution is roughly consistent with the observations, e.g. CR halo scale height is $\gtrsim 1$ kpc and the midplane energy density around $1\;{\rm eV/cm^{-3}}$ \citep{Ferr01,Gren15CRreview}. Our total $\gamma$ ray emission is also consistent with galaxies with similar gas surface densities \citep{Hopk19cr}. However, it is currently uncertain whether CR halos can extend and be dynamically important out to tens or even hundred {\rm kpc}, since observations of CRs and magnetic fields in those regimes are sparse.

While kinetic pressure is suppressed by CR feedback, the ratio between kinetic pressure and SFR is not reduced, likely because kinetic pressure is roughly fueled by stellar feedback and gas inflow, which correlate tightly with SFR. 

Note that our analysis includes pressure from the bulk flows so we only demonstrate {\it dynamical} balance in Hydro+ instead of {\it hydrostatic} balance, as also found in \cite{Gurv20prebal}. A steady inflow/outflow/fountain solution will still satisfy {\it dynamical} balance, but it will violate {\it hydrostatic} balance. We have shown in Fig. \ref{fig:pztur_ke} that the pressure from non-uniform motion (on $\sim 1 \;{\rm kpc}$ scale) dominates kinetic pressure around the mid-plane, but bulk flow pressure is more important above the plane. CR+ runs are close to hydrostatic balance. The gas is supported by kinetic pressure from non-uniform gas motion near the midplane and CR pressure above the disks.

In Appendix \ref{sec:mhdcr28}, we also examine the pressure balance in runs with CRs with a lower diffusion coefficient (CR+3e28). We find that they maintain dynamical balance, but the CR pressure in CR+3e28 drops much faster, so the kinetic and CR pressures are roughly in equipartition at even a great height.

\subsection{Vertical Gas Flows}
\label{sec:vergasflow}
\subsubsection{Multiphase gas flows}
We found in the last section that the sources of pressure support in Hydro+ and CR+ runs are different. We now explore whether CR+ and Hydro+ differ also in the gas kinematics and phase distributions.

\begin{figure}
\centering
 \includegraphics[width={0.45\textwidth}]{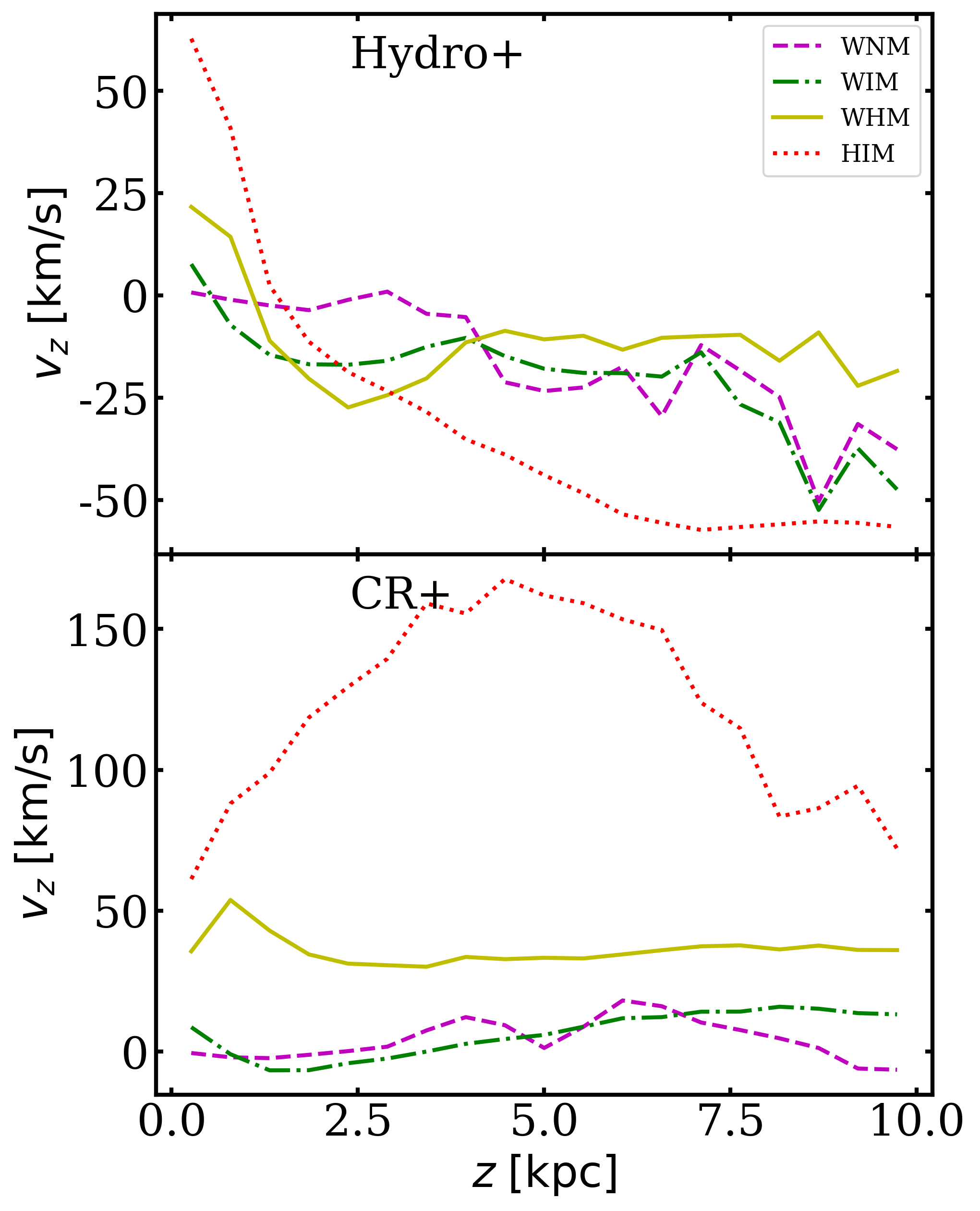}
\caption{Mass-weighted vertical velocities of different gas phases above {\bf m12i} galactic disk within $R=10\;{\rm kpc}$, averaged over the last 250 Myr. CRs untrap the hot gas near galactic disk, and also drive slow warm (ionized) gas outflows at high heights.}
\label{fig:gasdenvzz}
\end{figure}

\begin{figure}
\includegraphics[width={0.45\textwidth}]{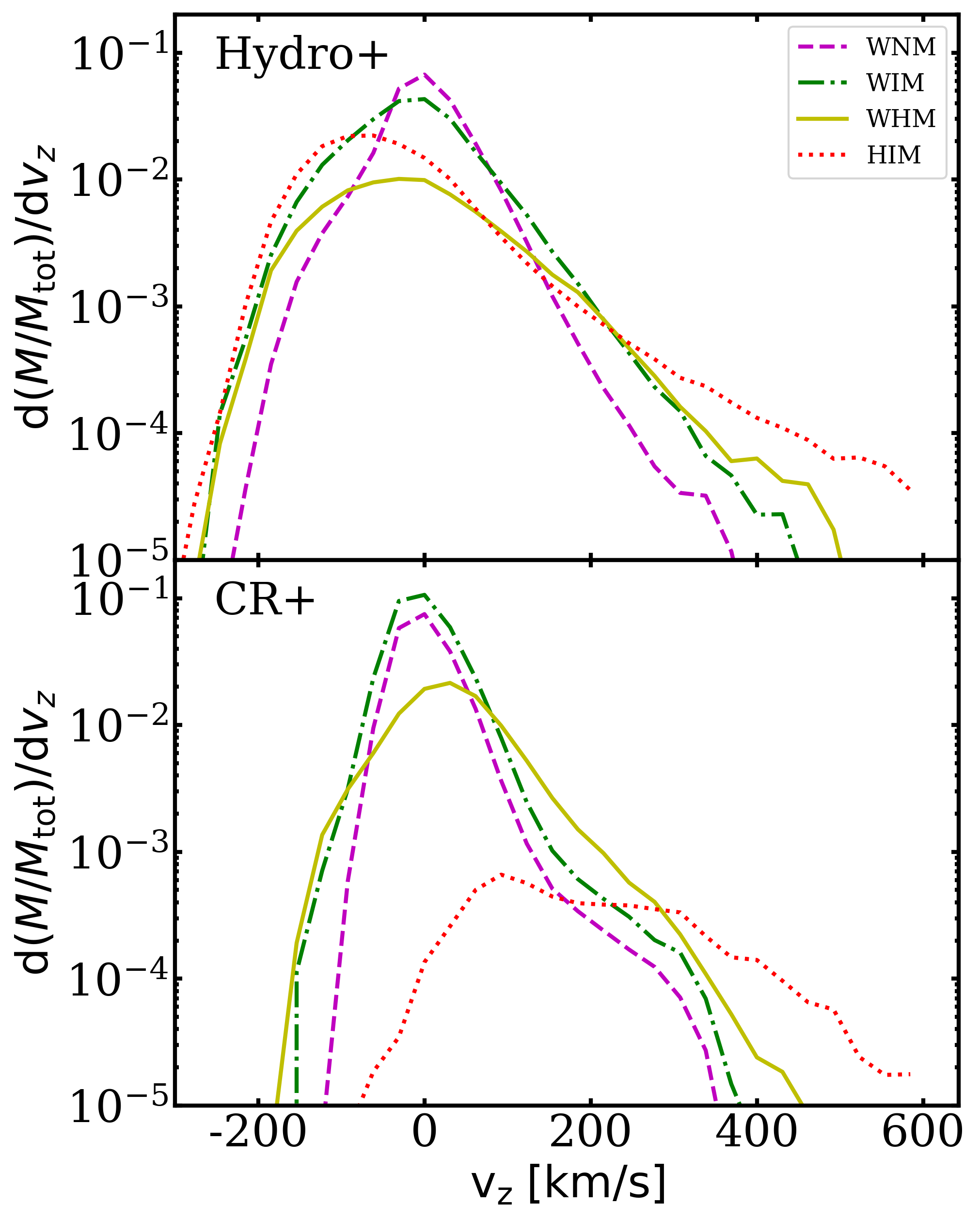}
\caption{The mass-weighted distribution of gas velocity in different phases in {\bf m12i} for $0\;{\rm kpc}<r<10\; {\rm kpc} $ and $2\;{\rm kpc}<z<10\;{\rm kpc}$, averaged the last 250 Myr. Results are similar for {\bf m12f} or {\bf m12b}.}
\label{fig:gasdenTvzm12ihighz}
\end{figure}

Fig. \ref{fig:slicem12tempvel} reveals the gas motions in the Hydro+ and CR+ runs, where arrows represent gas velocity and colors represent gas temperature. The most striking feature is the outflows from galaxies, as we also illustrated schematically in Fig. \ref{fig:schematic}. In Hydro+, there are some hot ($\sim 10^6{\rm K}$) and fast ($\sim 300\;{\rm km/s}$) superbubbles in the gaseous disks, some of which will break out from the disk, but most of the hot gas is not outflowing beyond 2 kpc, as we confirmed in Fig. \ref{fig:gasdenvzz}. In CR+, there are two types of outflows, the hot-fast ($\sim 10^6{\rm K}$; $\sim 300\;{\rm km/s}$) and warm-slow ($\gtrsim10^4{\rm K}$; $\lesssim 200\;{\rm km/s}$) types. The first type is still associated with SN bubbles, but the second type is likely wind driven by the CR pressure gradient (e.g. \citealt{Brei91,Sale14,Hopk20CRwind}).

There are also morphological differences between the outflows in Hydro+ and CR+. The hot fast outflowing gas in {\bf m12i} CR+ resemble galactic chimneys \citep{Norm89}:  the bubbles are elongated along the perpendicular direction within $z$ < 2 kpc, but then propagate radially beyond $z$ = 2 kpc. It is easier to see galactic chimney in CR+ than Hydro+, since hot gas more easily moves away from the disk along CR gradient, whereas in Hydro+ hot gas is confined by the ambient hot gaseous halos and gravity (see \S\ref{sec:outflowfate}). Compared to SN bubble driven winds, CR-driven winds are more uniform and moving more radially away from the galactic centers, since the CR distribution is roughly ellipsoidal around the galaxies. 

Another prominent difference between Hydro+ and CR+ is in the infalling gas, as we quantified in Figs. \ref{fig:gasdenvzz} and \ref{fig:gasdenTvzm12ihighz}. In Hydro+, the infalling gas is mostly $\sim 10^{5-6}{\rm K}$ with high relatively velocities ($\sim 50\;{\rm km/s}$) when high above the disk plane. There is a small amount of $10^{4}{\rm K}$ infalling gas, but it is mostly close to and associated with gaseous disks. On the other hand, with CRs, even the infalling gas is predominantly warm ($10^{4-5}{\rm K}$) and slow ($\sim 10\;{\rm km/s}$). In CR+, the hot ionized gas almost never falls to the disk due to its strong thermal plus CR pressure. 

We only show the net vertical outflows above the galactic disk, but the complementary analysis of FIRE-2 galaxies indicates that the bulk of galactic gas accretion actually proceeds largely along the disk plane at larger galactic radii \citep{Trap21gasinflow}. 

In general, Hydro+ runs have faster infalling and outflowing gas than CR+ (see Fig. \ref{fig:gasdenTvzm12ihighz}). The higher inflow rates in Hydro+ and thus the higher SFRs (Table \ref{tab:SIC}) cause stronger feedback and gas velocity in the Hydro+ runs. The weaker inflow in CR+ is likely a consequence of large-scale galactic outflows driven by CRs and the slowdown caused by non-thermal pressure gradients \citep{Ji20cr}. This results in lower SFR and explains why the kinetic pressure in CR+ is weaker than in Hydro+ in Fig. \ref{fig:prebal}. 

\subsubsection{The fate of hot/warm outflows}
\label{sec:outflowfate}
\begin{figure}
 \includegraphics[width={0.48\textwidth}]{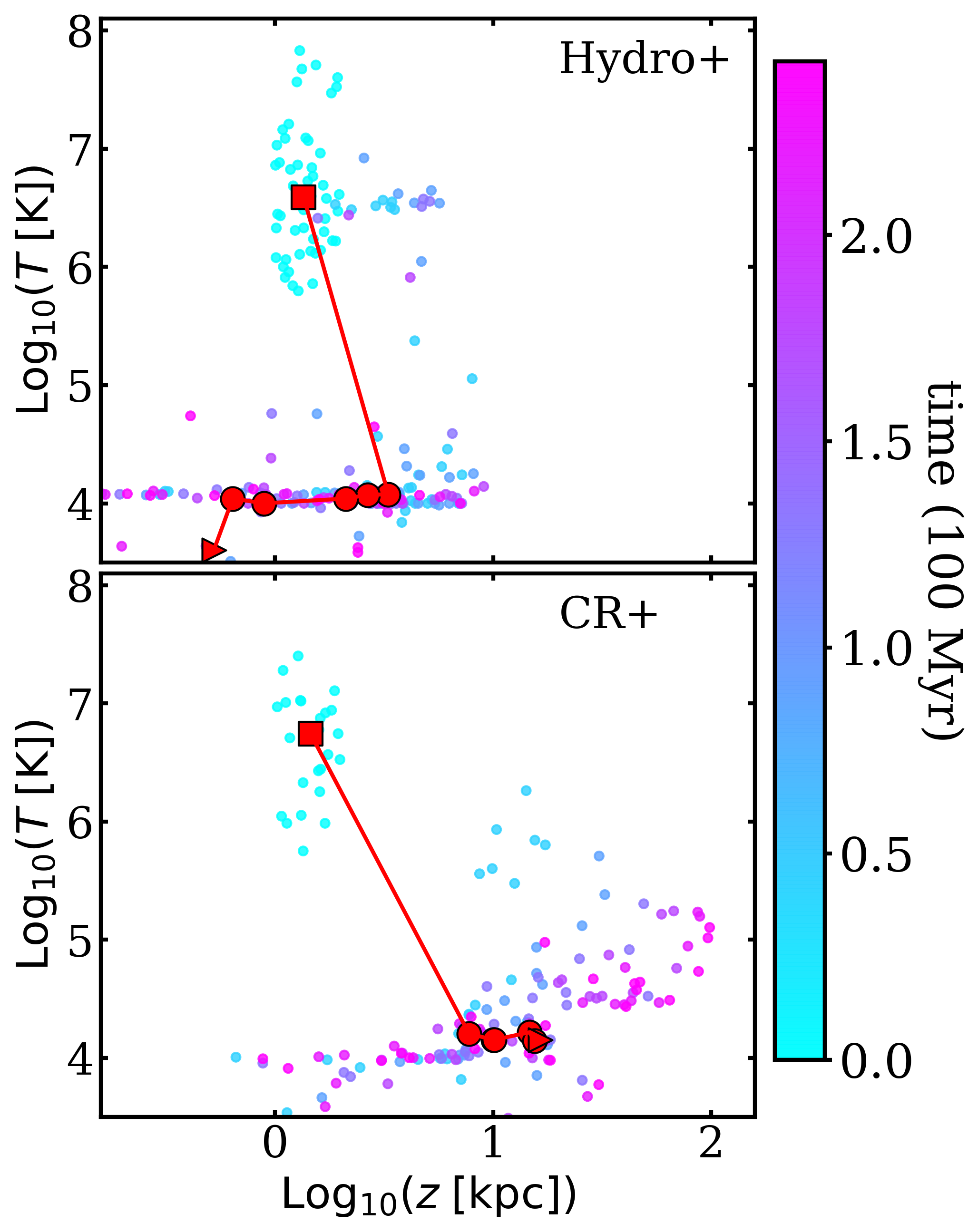}
\caption{ Mass averaged particle trajectories of hot fast outflowing gas near galaxies in {\bf m12i}. We follow particles initially (1) within $R$ < 10 kpc, 1 kpc < $z$ < 2 kpc, (2) with temperature $T$ > $5\times 10^5$ K, and (3) with vertical outflowing velocity $v_\mathrm{z}$ > 200 km/s at t=13.55 Gyr (red square). We then track their temperatures and heights above the disk ($z$) up to $t$ = 13.8 Gyr (red triangle). The red connected points show the median values for selected gas elements with a time separation of 44 Myr. Small circles represent individual gas particles whose colors indicate the times from the beginning. Hot outflowing gas is recycled in the Hydro+ runs, but not in the CR+ runs.}
\label{fig:Tztrackhot}
\end{figure}

The previous results seem to suggest that the fast moving gas from super-bubbles is trapped near galactic disk in Hydro+. Here we verify this by tracking fast hot gas particles near galactic disks in Fig. \ref{fig:Tztrackhot}. 

In Hydro+, initially hot and fast gas particles cool down to $10^4{\rm K}$ through adiabatic and radiative cooling, and return back to the midplane within $\sim$ 100 Myr. Gas particles are not able to reach the CGM ($>10\;{\rm kpc}$), and end up in galactic fountains. 

On the other hand, with the support of CRs, although hot fast outflows still cool down to $10^4~{\rm K}$, the bulk of this gas travels to a few tens of kpc above the disk and remain in the CGM for > 250 Myr. Thus, some of the warm and warm-hot gas in the CGM originates from hot galactic outflows. 

We have checked that a fraction of the gas even leaves the virial radius, contributing to the winds on scales of hundreds of kpc described in \cite{Hopk20CRwind}. Since only a tiny amount falls back to $\sim 1\;{\rm kpc}$, most of the hot fast outflows do not result in galactic fountains on short (< 250 Myr) time-scales. 

The dramatic difference between the fates of hot superbubbles can be understood when special properties of CRs are taken into account. \cite{Chan06CRconvect}  showed that (see also \citealt{Kemp20crstability}) if (1) CR pressure is significant\footnote{We have checked that the CR pressure is greater than the thermal pressure in most of the gas particles above the disk (with radii $R<10\;{\rm kpc}$ and heights $2\;{\rm kpc}<z<10\;{\rm kpc}$), except the hottest HIM and the most dense WIM.}, (2) CR diffusion is much faster than the dynamical time, and (3) CR pressure is decreasing outward, which is true if CRs are generated in the galaxy, then the gas is unstable for convection. {\it Hence, the gas in the CR-dominated regime is rising through the ``effective'' buoyancy.} On the other hand, the gas particles in Hydro+ cannot escape from the inner halo, partly because of the hot halo confinement. \cite{Hopk20CRwind} concluded that isotropic winds cannot escape above 10 kpc in a hot halo (without CRs), since the SN-driven winds do not have enough energy to plow through it. 

Even if we allow the winds to escape anisotropically (e.g. venting through holes in the gaseous halo), the hot halos can act against outflows by reducing their buoyancy, compared to warm halos (see similar phenomenon in \citealt{Bowe17entropywind}). Furthermore, we find that the fast hot outflowing gas particles cool too fast (in less than 40 Myr; Fig. \ref{fig:Tztrackhot}), which also limits the upward buoyancy. This might explain why the mass-loading factors of the MW-mass galaxies vanish in the FIRE simulations without CRs at low redshift \citep{Mura15,Mura17,Ster20CGM}\footnote{Even in Hydro+, occasionally intense star formation (${\rm SFR>30~\msun/yr}$) and the subsequent feedback can drive gas to $\sim 100\;{\rm kpc}$  \citep{Pand21outflow}, e.g. during mergers in m12f, but such events are rare in normal $L_\star$ galaxies.}. 

CR pressure gradient also reduces the required velocity to escape from the inner halo.  Unlike gas thermal pressure which mostly follows gas particles, CRs can diffuse rapidly, so CR pressure gradient acts like a background to counter-balance a significant fraction of the gravitational force. Thus the gas particles at the midplane could reach $z=10\;{\rm kpc}$ with $v_z\sim 100\;{\rm km/s}$, a factor of two smaller than the required velocity without CRs.

\begin{figure}
 \includegraphics[width={0.45\textwidth}]{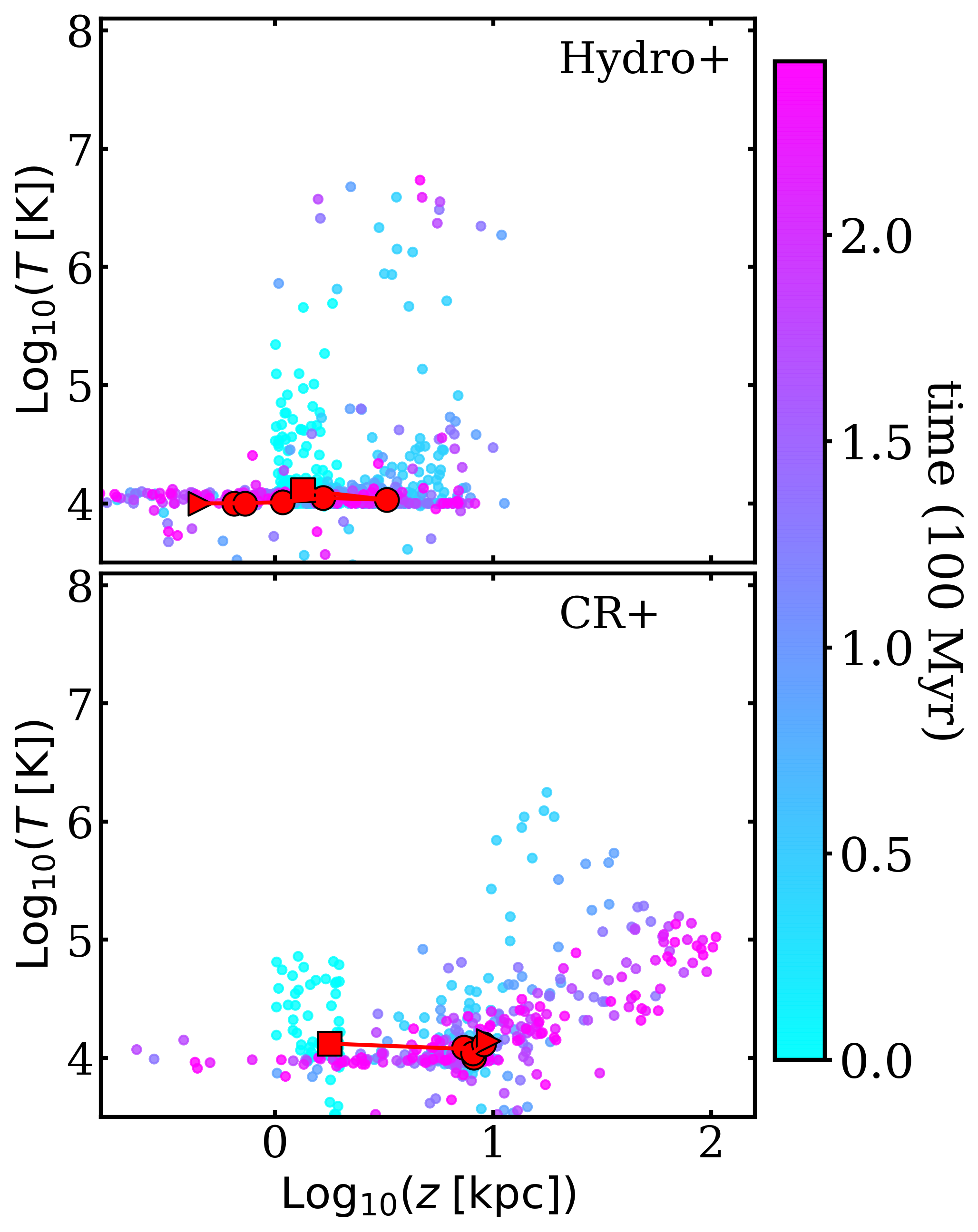}
\caption{Mass averaged particle trajectories of warm and warm-hot fast outflowing gas near galaxies in the temperature-height space in {\bf m12i}. We follow particles (1) within $R$ < 10 kpc, 1 kpc < $z$ < 2 kpc, (2) with temperature between $10^3$ K and $5\times 10^5$ K, and (3) with vertical outflowing velocity $v_\mathrm{z}$ > 200 km/s at $t$ = 13.55 Gyr (red square). Then we track their temperatures and heights up to $t$ = 13.8 Gyr (red triangle). The red connected points show the median values with a time separation of 44 Myr. Small circles represent individual gas particles whose color indicates the times from the beginning of our particle tracking.}
\label{fig:Tztrackwarm}
\end{figure}

Finally, we investigate the fate of the fast warm gas outflow in Fig. \ref{fig:Tztrackwarm}, which shows that, in CR+, this gas is able to escape from the inner halo, some of which is able to reach $\gtrsim$ 100 kpc, contributing to the warm CGM. In Hydro+, fast warm or hot gas particles are trapped within 10 kpc from the galaxies, so they constitute classical or warm galactic fountain flows \citep{Houc90warmgf}.

\subsection{Distribution of multiphase gas}
\label{sec:verticaldensity}

\begin{figure*}
\includegraphics[width={0.45\textwidth}]{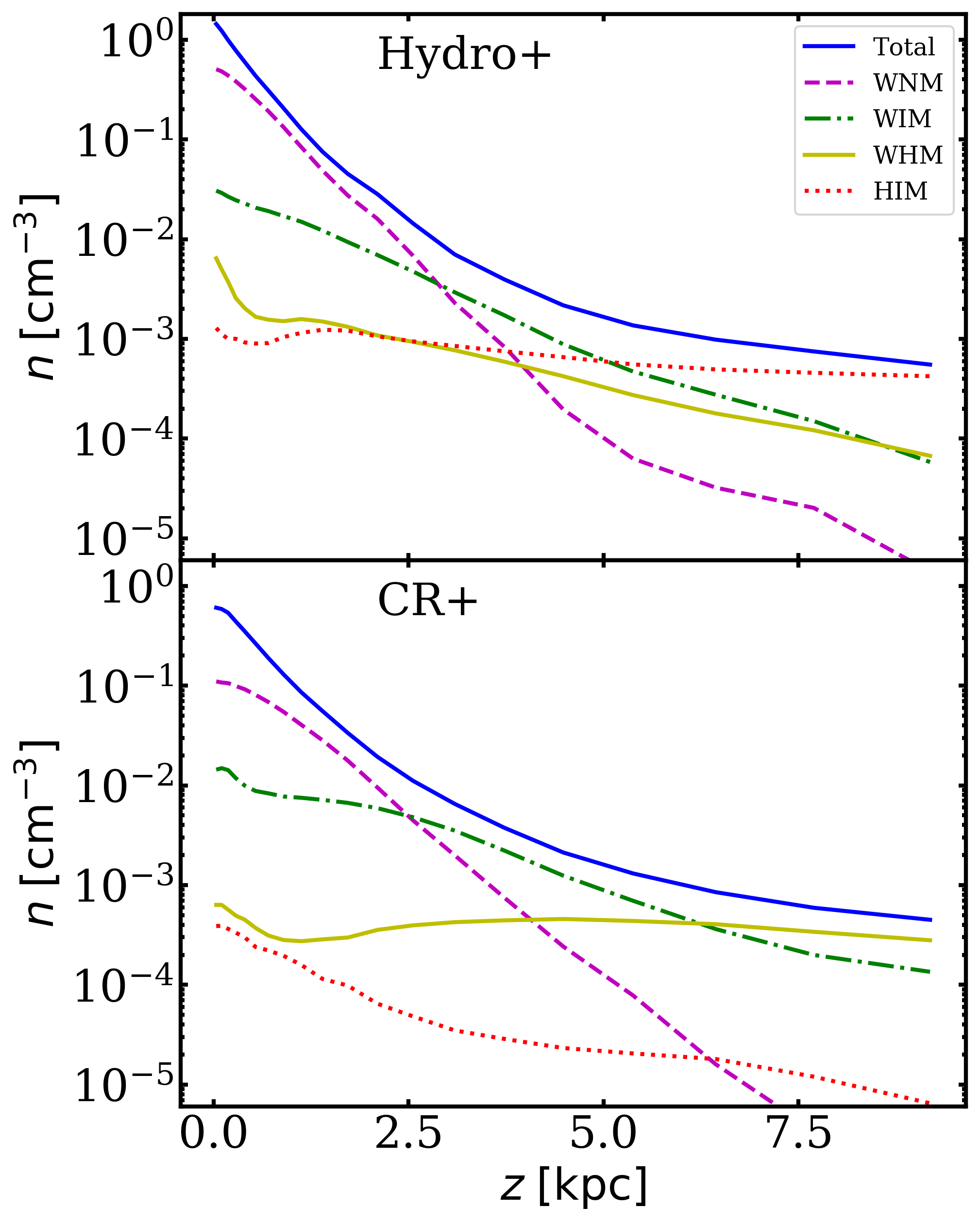}
\includegraphics[width={0.45\textwidth}]{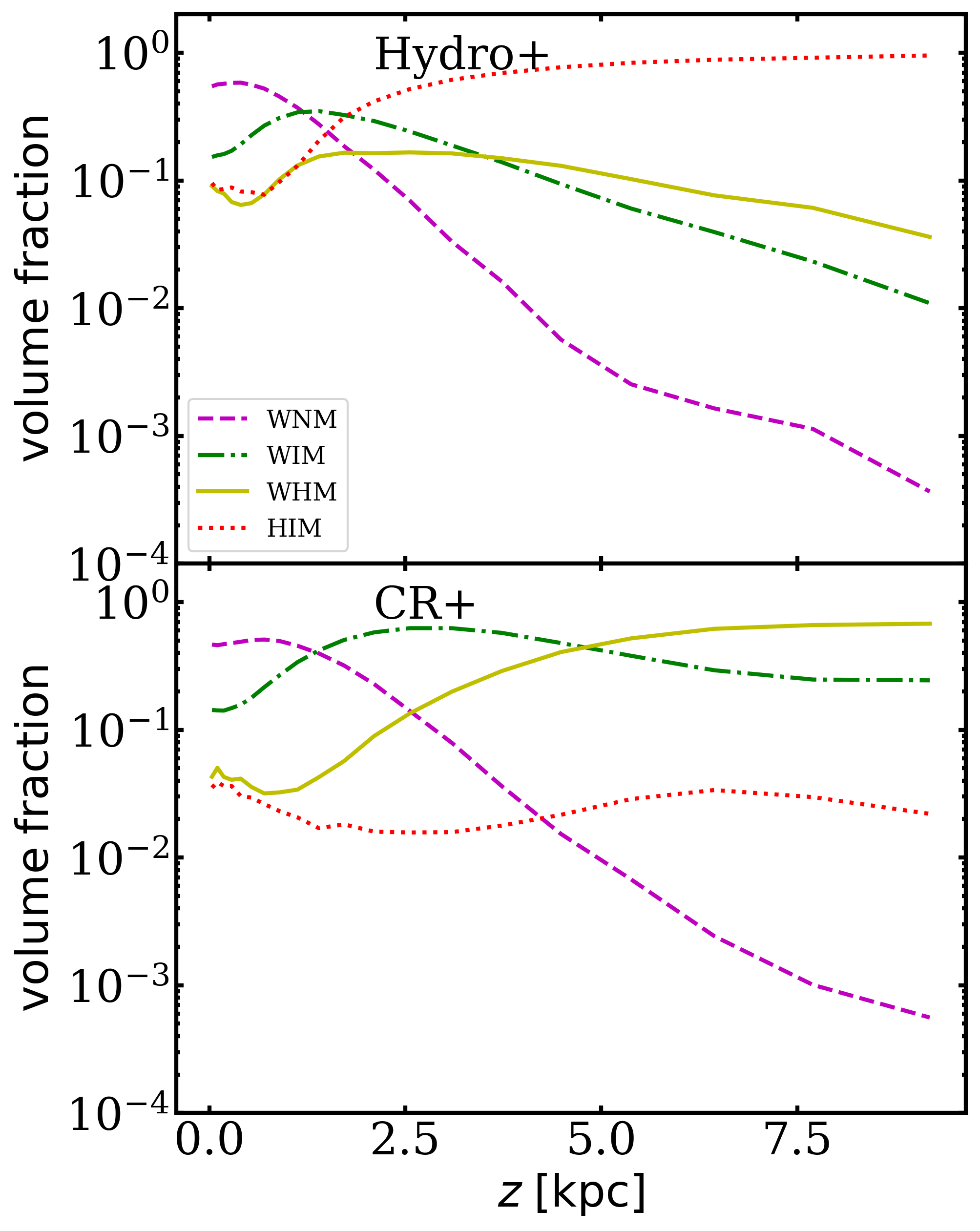}
\caption{Gas densities ({\it left}) and volume filling fraction ({\it right}) in different phases as a function of height in {\bf m12i}, averaged over 250 Myr. The phases are defined in \S\ref{sec:multiphasedef} and ``Total'' represents the total gas density. Hydro+ has a higher fraction of hot gas (HIM), especially at greater heights CR+ has more extended warm ionized gas (WIM) and warm hot gas (WHM). CRs can help support more gas with a smaller SFR. At a few kpc above the plane, most of the volume is occupied by the hot gas in Hydro+, whereas the WIM and WHM occupies most of the volume in CR+.}
\label{fig:gasdenTz}
\end{figure*}

Fig. \ref{fig:gasdenTz} shows the vertical distribution of volume-weighted gas density $n$, defined as $\langle \rho\rangle/m_{\rm H}$.
CRs have a strong impact on the vertical distribution of gas phases. The total vertical gas density distribution in Hydro+ is slightly more concentrated than in CR+. Hydro+ has midplane density two to three times larger than CR+ (as expected from higher SFRs in Hydro+ simulations). Due to their similar total pressure support, CR+ and Hydro+ have similar gas densities at a large height. In CR+, the warm and warm-hot phase are more extended and contain more mass at large $z$, while the hot gas is more concentrated.
 
\begin{figure}
 \includegraphics[width={0.5\textwidth}]{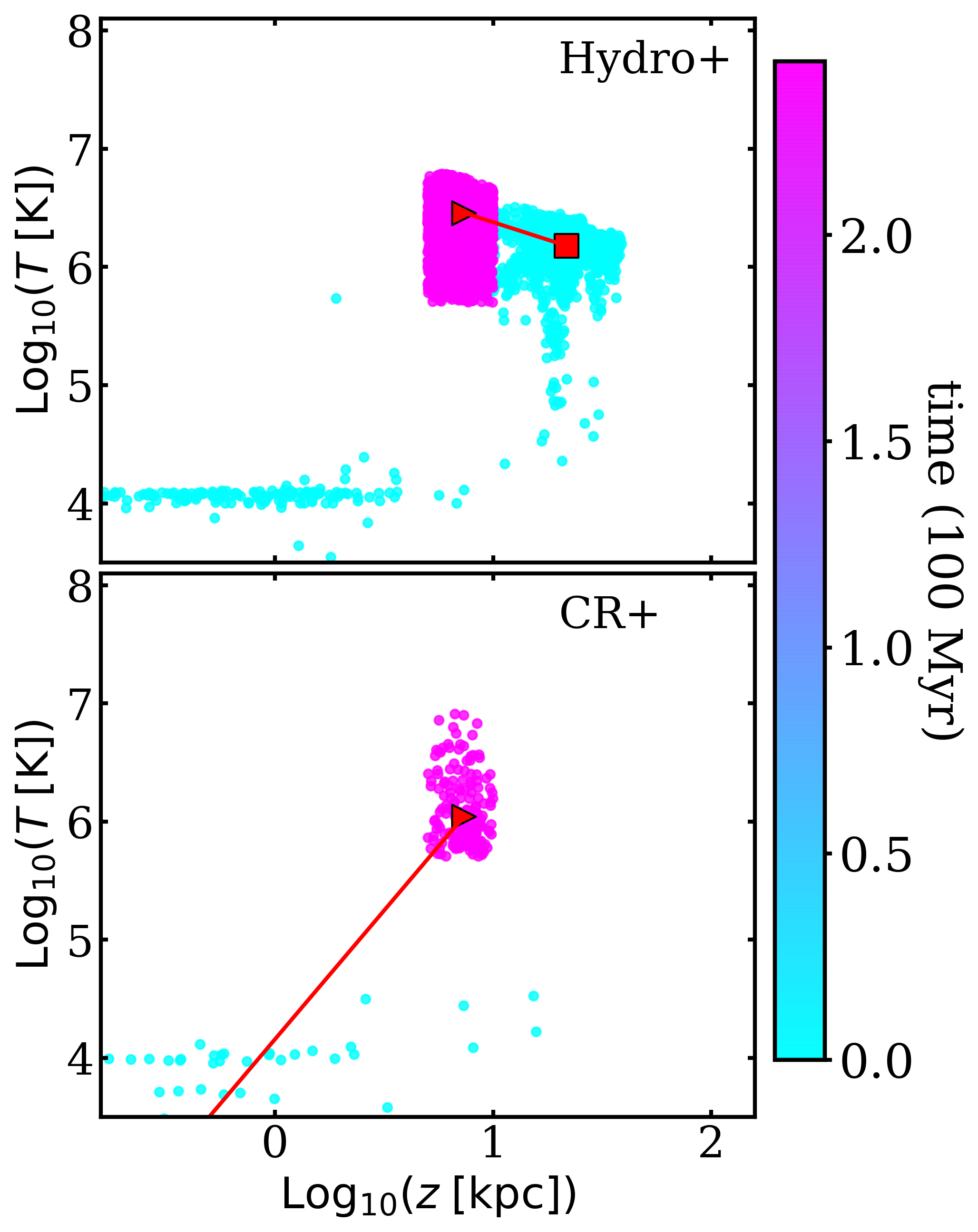}
\caption{ Temperature-height evolution of the hot gas in the inner halo.
We select the hot gas ($T>5\times 10^5\;{\rm K}$) particles in {\bf m12i} within $R$ < 10 kpc and 5 kpc < $z$ < 10 kpc at $t$ = 13.8 Gyr ({\it dark blue}) and track those particles back to $t$ = 13.55 Gyr ({\it light blue}). Also shown are mass weighted average temperature and height at $t$ = 13.8 Gyr (the present; {\it red triangle}) and $t$ = 13.55 Gyr ({\it red square}). 
In Hydro+, a significant amount of hot gas comes the CGM further out in the halo, but most of the hot gas in CR+ is from the galaxy.}
\label{fig:Tztrack}
\vspace{-1.5em}
\end{figure}

There are two major reasons for the difference in the gas distribution. First, the gas density distribution, especially at a large height, depends on the influence of the CGM. As shown in \cite{Ji19CRCGM}, the CGM is predominantly hot $\sim 10^6\;{\rm K}$ without CRs. The HIM density in Hydro+ has much flatter profile because the CGM close to the galaxy ($\lesssim 10 {\rm kpc}$) is supplied by the hot gas that originates further out (and moves inward in a ``cooling-flow'' fashion) as we have verified in Fig. \ref{fig:Tztrack}. In CR+, \cite{Ji19CRCGM} found the gas in the CGM is mostly around $10^5{\rm K}$ with little hot component, so the WHM gas distribution is extended, but the hot gas distribution is more concentrated.

Second, CRs can support the WHM though CR pressure and heat the gas through turbulent heating shown in \cite{Ji20cr}. CR hadronic and streaming heating can also provide extra heating to the WHM \citep{Wien13CRWIM,Vand18CRWIM}, although it is less important in our runs \citep{Ji19CRCGM}. 

Consequently, the WHM is the most dominant gas phase at a large height in CR+, while the HIM is the most dominant at a large height in Hydro+ \footnote{Note that in ISM observations, the WHM with $\gtrsim2\times10^5\;{\rm K}$ gas is included in the ``coronal'' or ``hot'' component (e.g. \citealt{Jenk78OVI}). We should account for this when comparing with the ISM studies, e.g. the volume filling in the ISM.}.

Finally, we also show the volume fractions of different gas phase in Fig. \ref{fig:gasdenTz}. In Hydro+, hot gas fills most of the volume, whereas in CR+, the WIM and WNM phases fill the largest volume fraction. CRs not only suppress the volume of hot gas, they also boosts the WIM and WHM volumes by supporting them in the disk-halo interface.

In Appendix \ref{sec:mhdcr28}, we find that the CR diffusion coefficient has large effect on the vertical gas distribution. A lower CR diffusion coefficient (CR+3e28) leads to a lower CR pressure away from the disk and less extended CR halo,  Consequently, the kinetic pressure plays a bigger role. While in this model, CRs still supports a lot of warm gas a few kpc away from the disk, hot gas is more extended than that in the default CR+. However, this low diffusion coefficient is disfavored by the observed $\gamma$-ray emission from nearby galaxies \citep{Chan18cr,Hopk19cr}.

\subsection{The Effects of Cosmic Rays on Midplane Velocity Dispersion}
\label{sec:midturSFR}
\begin{figure}
 \includegraphics[scale=0.5]{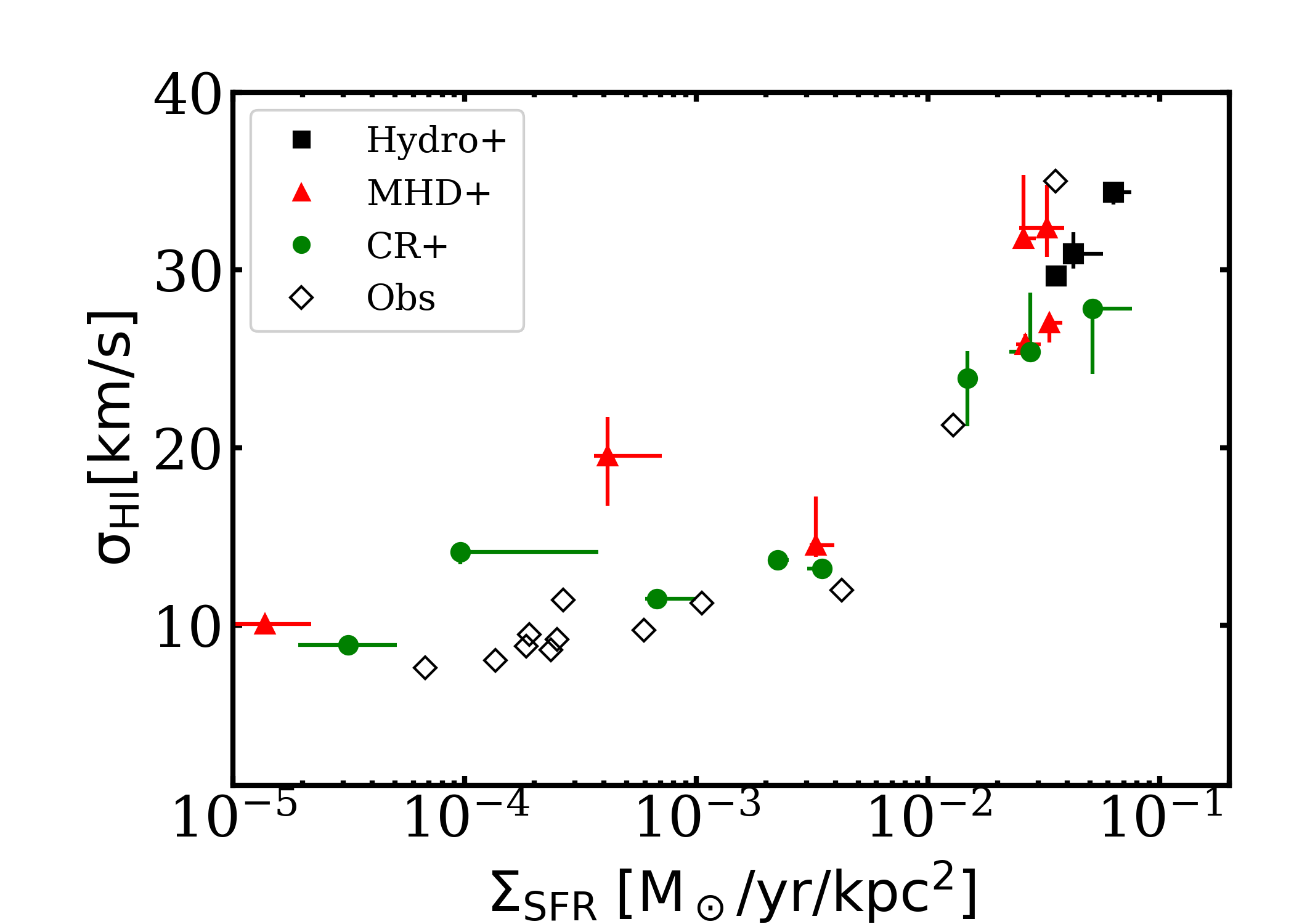}
\caption{Mass-weighted HI velocity dispersion ($\sigma_{\rm HI}$; including both turbulence and thermal broadening), averaged over last 250 Myr, as a function of star formation rate (averaged over 10 Myr) per unit area, $\pi (3r_s)^2$, where $r_s$ is stellar radial scale length), $\Sigma_{\rm SFR}$. We show results for a compilation of Hydro+, MHD+ and CR+ runs (sees the main text for details). CRs do not change the relationship between $\sigma_{\rm HI}$ and $\Sigma_{\rm SFR}$. }
\label{fig:speedmwsfr}
\end{figure}

In \S\ref{sec:hydrobal}, we found that kinetic pressure from velocity dispersion dominates the midplane pressure. Although CR pressure is not important at the mid-plane owing to its large scale height (Fig. \ref{fig:prebal}), the question remains if CRs can affect the velocity dispersion in the midplane indirectly (e.g. by suppressing or enabling winds or inflows).

Fig. \ref{fig:speedmwsfr} addresses this question by comparing the total velocity dispersion (kinetic plus thermal) of the HI gas, $\sigma_{\rm HI}$\footnote{We estimate HI by subtracting molecular hydrogen fraction, ${\rm H}_2$, calculated using the \cite{Krum11} approximation based on local metallicity and gas column density. We applied the Sobolev approximation and take the gas kernel length as the local density scale height. Finally, we measure the mass-weighted velocity dispersion (in vertical direction) within +/-250pc of the disk plane.},
 as a function of SFR per unit area in simulations and observations\footnote{For a pixel by pixel comparison of the FIRE simulation (without CRs) with spectrally-resolved observations of the ISM gas velocity dispersion, see \cite{Orr20veldis}.}. In order to compare with the observations at lower SFRs, we have also included additional runs from \cite{Hopk19cr}, expanding our sample to include {\bf m11g}, {\bf m11d}, {\bf m11h}, {\bf m11b}, {\bf m11f}, {\bf m12i}, {\bf m12f}, and {\bf m12b} simulations.

 The observational data is obtained from \cite{Dib06} and references therein\footnote{Note that the rightmost observation point is from the inner region of NGC 2915, while all other points are averaged for whole galaxies.}. We find a good agreement between simulations and observations. Velocity dispersion is constant at low SFRs (usually dwarf galaxies), because most of the velocity dispersion is from thermal broadening of $\lesssim 10^4{\rm K}$ HI gas. This is not affected by CR feedback, as CR pressure is not significant in the ISM of dwarf galaxies due to the rapid escape of CRs at low density \citep{Lope18SMCCR}.
 
 In $L_\star$ galaxy runs, mid-plane turbulence increases with $\Sigma_{\rm SFR}$ with a steepening at high $\Sigma_{\rm SFR}$. The CR+ runs have a lower $\sigma_{\rm HI}$ and $\Sigma_{\rm SFR}$ {\it along} the observed relation, confirming that CRs reduce velocity dispersion indirectly through suppressing inflow rates, SFRs, and/or feedback.

%% file: discussions.tex
\section{Discussion}
\label{sec:discussion}

\subsection{Comparison with other theoretical studies}

\subsubsection{Dynamical balance}

We showed that with or without CRs, the bulk flow pressure is not negligible a few kpc above the galactic disk (Figs. \ref{fig:prebal} and \ref{fig:pztur_ke}), so {\it hydrostatic balance} (balance without bulk flow pressure) is violated. This is in disagreement with the early theoretical studies, e.g. \cite{Boul90} and \cite{Ferr98hot}. Our finding is more inline with the recent theoretical studies by \cite{Boet16DIGbalance,Boet19DIGbalance} which found that hydrostatic balance does not hold in some spiral galaxies.

In our simulations, {\it dynamical balance} (that includes also the bulk flow pressure) is approximately satisfied: for $z\gtrsim 1\;{\rm kpc}$, the deviations are typically less than 10\% (see also \citealt{Gurv20prebal}). These findings are in qualitative agreement with previous vertical balance studies using a variety of codes and assumptions while neglecting CRs, e.g. \cite{Hill12}, \cite{Kim15verequil}, \cite{Vija20verbal}, \cite{Beni16diskprebal}, and \cite{Gurv20prebal}. Similar to our findings, these studies found that kinetic pressure dominates near the mid-plane while thermal (plus kinetic) pressure dominates a few kpc above the midplane. We also agree with \cite{Hill12}, \cite{Kim15verequil}, and \cite{Vija20verbal} that the magnetic tension is not important in supporting the gas above the disk. 

When CRs are included, we found that CR pressure dominates well above disks while kinetic pressure dominates around the mid-plane. This is in contrast to \cite{Giri18} who found CR pressure is most important even at the midplane, likely because of their weaker thermal feedback and slower CR diffusion (see the discussion in \citealt{Kim18solarneigh}). 

By post-processing high resolution ISM simulations with CRs, \cite{Armi22CRTvary} also found that the CR pressure gradient is dominant for $|z|>0.5\;{\rm kpc}$, while the CR pressure gradient is insignificant for $|z|<0.5\;{\rm kpc}$. This agrees qualitatively with our findings -- CRs are important in supporting gas and driving winds in the disk-halo interface ($|z|>0.5\;{\rm kpc}$).

\subsubsection{Vertical gas structure and flows}
Our Hydro+ and CR+ runs have extended gas distribution; e.g. warm gas with scale heights $\gtrsim$ {\rm kpc}, in agreement with the observation. This differs from the results from the local or stratified box simulations with small vertical extents \citep{Wood10,Kim15verequil}. 

There are several reasons for the difference. First and most importantly, a simulation domain with a sufficient height ($\gtrsim 10\; {\rm kpc}$) is required to capture galactic fountain flows and allow extended vertical gas distributions \citep{Hill12,Vija20verbal}. Second, our simulations are cosmological, so they include the CGM, which can increase the scale height (especially for the hot gas; see Fig. \ref{fig:Tztrack}). Finally, local or vertical stratified boxes cannot capture the correct outflow boundary conditions \citep{Mart16}. 

In our simulations, CR pressure boosts gas scale height. But the boost is modest compared to \cite{Giri16CR}, who found that CR feedback extends gas scale height by more than an order of magnitude, likely because their thermal feedback is too weak (see the previous section) and because they assume a much lower diffusion coefficient that traps most of the CR energy density in the disk.

Our simulations without CRs (Hydro+) are consistent with other studies finding that the warm-hot ($T\sim 2\times 10^4-5\times 10^5{\rm K}$) gas is not substantial (owing to thermal instability) and the hot gas dominates at a large height \citep{Hill12,Vija20verbal}. In contrast, CR feedback can increase the amount of warm-hot gas, and reduce the amount hot gas away from the galaxies, as also noted previously in, e.g., \cite{Boot13} and \cite{Sale16cgm}.

CR-driven warm galactic winds have already been found in several other studies (e.g. \citealt{Boot13,Giri18}). Our study differs from them by focusing on how thermal and CR pressures can drive winds in coordination in cosmological galaxy formation simulations. Our simulations include both warm CR-driven and hot thermal/super-bubble driven winds.

Hybrid thermal-CR winds were studied in previous 1D analytic, idealized, models \citep{Brei91,Brei99CRwindhotbubble,Ever08}. However our work is unique in the following aspects:

First, our CR propagation models are distinct from previous 1D models. We consider streaming and constant diffusion coefficient everywhere calibrated to match the $\gamma$-ray emission, whereas most 1D models assume Alfv{\'e}nic streaming above galactic disks. However, regardless of the propagation models, both our simulations and these 1D models predict CRs can help drive winds in MW-mass galaxies.

Second, we are able to simulate and quantify the combination of hot super-bubble-driven winds and CR-driven warm winds in $L\star$ galaxies in 3D cosmological settings. Cosmological setting and more complete physics enable us to also show how CR feedback can suppress galactic fountains or, at least, significantly increase their return periods by preventing the cooled winds from falling back (\S\ref{sec:outflowfate}).

\subsection{Comparison with observations}
\label{sec:comobs}
\subsubsection{Hot ionized gas}
\label{sec:HIMSFR}
\begin{figure}
\includegraphics[width={0.48\textwidth}]{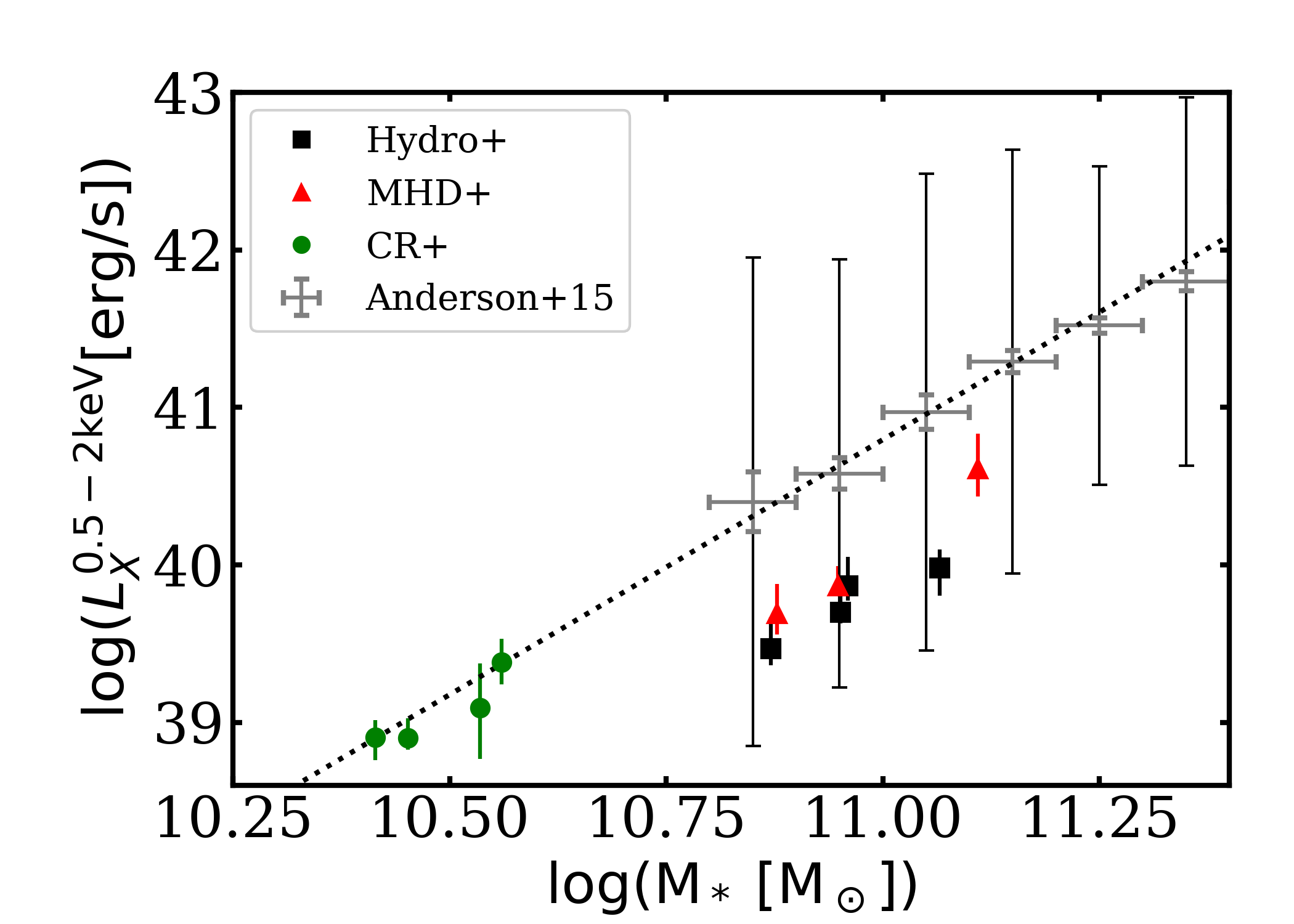}
\includegraphics[width={0.48\textwidth}]{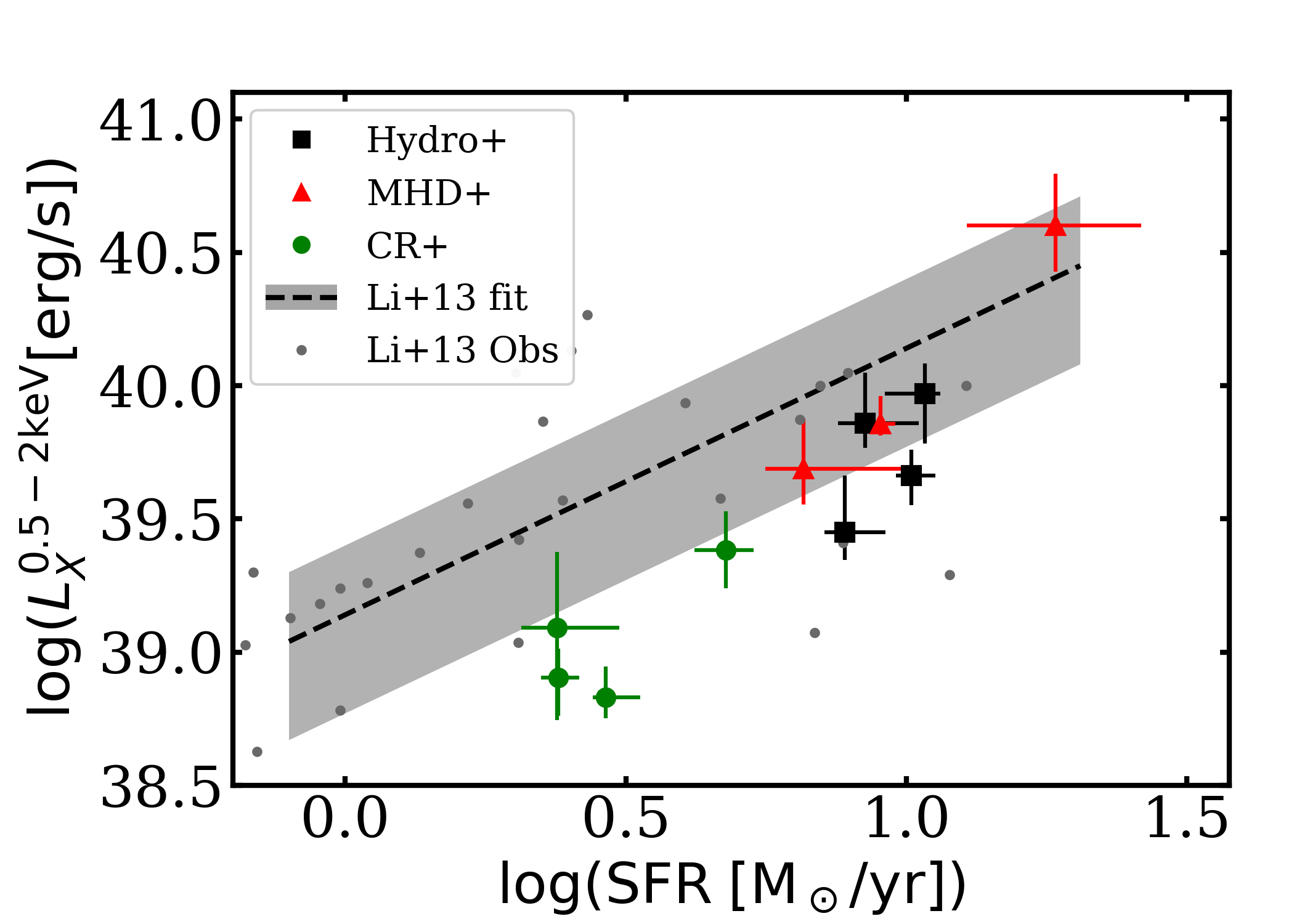}
\caption{{\bf Top:} soft X-ray luminosity (0.5-2 keV) as a function of stellar mass. The colored error-bars represent the soft X ray luminosity within 200 kpc of our simulated galaxies (avaraged over the last 250 Myr). The grey errorbars represent the observation data of the total soft X-ray luminosity from \protect\cite{Ande15xraystack}. The thick errorbars show the measurement errors, while the thin error-bars show the upper limit of the instrinsic scatter. The dotted line is the best power-law fit to the observation points. {\bf Bottom}: soft X-ray luminosity as a function of star formation rate. The colored error-bars represent the soft X-ray luminosity within 20 kpc and SFR over 10-Myr interval of our simulated galaxies. The dashed line and grey region represent the best linear fit and corresponding scatter of the observations of nearby galaxies from \protect\citep{Li13xraysfr}. Individual observation data are also over-plotted as gray points.}
\label{fig:sfrhotgasmass}
\end{figure}

Hot ($>5\times10^5{\rm K}$) gaseous halos have been detected in the MW and many external spiral galaxies from their soft X-ray emission \citep{Snow98hotxray,Henl13hotxray} and absorption of high ions  \citep{Gupt12hothalo}. Recently, \cite{Kaar20MWxrayhot} observed the soft X ray emission in the MW and inferred that the midplane hot gas density around solar circle is around $10^{-3}{\rm cm}^{-3}$ and its scale height is around 2 kpc (see also \citealt{Hagi10xray}). This scale height is slightly larger than that in our CR+ runs, but smaller than our Hydro+ runs (Fig. \ref{fig:gasdenTz}). 

Hot halos can be extended to 50-80 kpc around some spiral galaxies, e.g. \cite{Ande16hothalospiral,Dai12hothalospiral}, which could be at odds with our concentrated hot gas halo in CR+. However, to the best of our knowledge, these extended hot halos are detected only around galaxies much more massive than our simulated galaxies, objects with the AGNs, or in starbursts \citep{Wang01hot,Frat02xrayngc2403,Das20spiralhot}, which are not included in our simulation sample. 

For less massive spiral galaxies, with masses closer to our simulated ones, soft X-ray emitting halos halos are not observed to tens of kpc \citep{Tyle04xrayspiral,Tull06hotgaslatetype}, although the low density hot gas tens of kpc away from these galaxies is likely below current X-ray observational limits. 

We compare our simulations to diffuse soft X-ray observation in Fig. \ref{fig:sfrhotgasmass}. We considered all of the available {\bf m12} galaxies here, including {\bf m12m}, to show X-ray in more massive or starburst galaxies\footnote{For a comparison with the X-ray observation in the FIRE-1 galaxies without CRs, see \cite{vand16hot}.}. We calculate the soft X-ray emission of individual gas particles by interpolating a table computed with the Astrophysical Plasma Emission Code ({\small APEC}, v3.0.9; \citealt{Smit01APEC,Fost12APEC}), which assumes the gas is optically thin and in collisional equilibrium\footnote{We neglect absorption from HI in the calculation of X-ray luminosity; observations typically correct for this effect and often found it to be negligible \citep{Li13xraysfr}.} 
Then we sum over all of the X-ray luminosities from diffuse emission of hot gas within a given (3D) radius.

In the top panel of Fig.\ref{fig:sfrhotgasmass}, we compare our simulation against the stacked soft X-ray observation from \cite{Ande15xraystack}\footnote{We do not include observations lower than $\log(M_*/\msun)=10.8$ due to the contamination of low mass X-ray binaries, but the contamination is small above that mass (see \citealt{Ande15xraystack} for the estimate of the contributions from X-ray binaries). }. The thin grey errorbars represent the intrinsic scatters, which come from some unknown contribution from X-ray binaries and low luminosity AGN activities. Hydro+ and MHD+ runs lie slightly below the observation points, but still within the estimated range. Note that the Hydro+ and MHD+ runs are slightly above the observed stellar-to-halo mass relation \citep{Hopk19cr}, so their X-ray luminosity is likely even a better match for a given halo mass. The CR+ runs agree well with the extrapolation of the observational points.

In the lower panel of Fig. \ref{fig:sfrhotgasmass}, we show that the amount of hot gas is roughly proportional to SFRs for all Hydro+, MHD+, and CR+ runs (see also \citealt{Tull06hotgaslatetype}). Hydro+ and MHD+ broadly agree with the observation data listed in \cite{Li13xraysfr}. Although the CR+ runs have weaker X-ray emission, at the same time they have lower SFRs, so they still lie close to the low end of the observed luminosities. 

Another observational constraint on the hot gas is the Sunyaev-Zel’dovich  (S-Z) effect \citep{Suny70SZeffect}. However, current observations \citep{Plan13SZ,Scha20ACT} are limited to halos more massive than $M_{\rm halo}\sim 10^{12.5}\msun$, and/or exclude the central regions \citep{Breg21SZMW}. Thus, these results are not yet able to probe the disk-halo interface of $L\star$ galaxies.

In the CGM, the low and intermediate ion absorbers are in a very good agreement between CR+ and observations \citep{Ji19CRCGM}. But there may be some tension in the high ions (e.g. NeVIII), although the observation data is sparse. Future comparisons with high ion absorption  could provide extra constraints.

\subsubsection{Warm gas distribution and flows}
\label{sec:obswarm}

At the solar circle, the observed HI scale height is around 200 pc \citep{Dick90HIrev} and WIM scale height around 1-2 kpc \citep{Reyn91HII,Gaen08WIM}. A number of external spiral galaxies also keep a fraction of their HI gas in a thick layer (with thickness a few kpc) above the midplane \citep{Swat97HIrot,Frat01EPHI}. The WIM is also observed in many edge-on spiral galaxies \citep{Hoop99DIG,Rand96DIG,Rand00extendedDIG}. with very extended H$\alpha$ emissions (up to 5 kpc in some cases). These HI and WIM scale heights are in general agreement with our simulations with or without CRs.

There are some signatures of WIM outflows \citep{Voig13DIGoutflows} in moderately star forming galaxies, which could be driven by CRs. For example, using the observational data of nearby radio halos and 1D CR propagation models, \cite{Hees18obsCRwind} and \cite{Schm19crwind} found evidence of CR-driven disk winds.

\subsubsection{Warm-hot (transitional) gas }
\label{sec:obsWHM}
The WHM is observed in the galactic corona with a $\sim$3 kpc scale height around the MW \citep{Putm12GGH,Sava09transgas} and 8 kpc in an edge-on spiral galaxy (NGC 4631)\citep{Otte03spiralOVI}. Observations of quasar absorption sightlines suggest that the WHM can extend to $\gtrsim 100\;{\rm kpc}$, which may be responsible for a significant fraction of the missing baryons (see, e.g. \citealt{Tuml11LstarOVI,Werk13,Lehn20M31CGM}).

However, it is challenging to maintain the warm-hot ($\sim 2\times10^4-5\times10^5{\rm K}$) gas away from galaxies without CRs, since the WHM is thermally unstable. For example, Fig. \ref{fig:gasdenTz} shows that our Hydro+ runs have less extended WHM than CR+ runs. In a more quantitative study, \cite{Ji20cr} found the distribution of the OVI absorption in the CGM in the CR+ runs agree well with observations, while our Hydro+ runs underestimate the absorption strength of OVI.

\subsection{Caveats}
\label{sec:caveats}
The resolution of our simulations is not sufficient to directly resolve some important processes, such as (a) the cooling radii of SN explosions \citep{Hopk18sne,Gutc20LYRA}, 
(b) instabilities that disrupt small warm clouds moving through hot medium \citep{Cowi77cloudcrush,Klei94clcrush}, (c) physical processes that prevent the disruptions of the clouds \citep{McCo15Bclcrush,Scan15}, (d) the sub-pc thermal instabilities \citep{McCo18shattering,Spar3Dshattering} (e) the thermal instabilities enhanced by weak magnetic fields \citep{Ji18}. 
Thus, the precise multiphase gas distribution, especially in Hydro+ or MHD+, is still uncertain. 

Interestingly, consequences of some of these processes can be less severe with significant CR pressure, e.g. thermal instabilities might not results in poorly resolved, dense clouds \citep{Buts20CRstability} and clouds can be accelerated without disruption \citep{Wien17CRcoldclouds,Bust21crism}.

However, the microphysics of CR transport still remains order-of-magnitude uncertain. We parameterize this here with a simple streaming+constant diffusivity model, with the constant diffusion coefficient calibrated to reproduce $\gamma$-ray and grammage observations. Realistically, CRs interact locally with Alfv{\'e}n waves and thermal gas, resulting in variable streaming speeds and diffusion coefficients in different ISM phases (e.g. \citealt{Zwei13,Thom21TMCRAREPO,Armi21cr,Thom22CRvary}).

In particular, \citet{Hopk20CRtrans,Hopk20CRtransgalaxy} showed that a variety of models with more complicated dependence of CR transport on local plasma properties \citep[see e.g.][]{Zwei13} can reproduce the same observables close to the disk (primarily coming from the ISM of the MW and nearby galaxies) while behaving differently in the CGM. These models generally gave rise to faster CR escape outside the inner halo. Hence, CR pressure profiles drop more rapidly, so the results are ``in-between'' our CR+ and MHD+ runs here. Constraining which (if any) of these best represents reality remains an important question for the future work.

Since the CR propagation physics is uncertain away from galaxy disks (\S\ref{sec:caveats}), future observations of extended synchrotron emission (e.g. \citealt{vanH13LOFAR}) and $\gamma$-ray emission of nearby galaxies can provide valuable constraints (e.g. \citealt{Karw19M31gamma}).

Another related concern is the effective diffusion coefficient (determined by the observed hadronic $\gamma$-ray luminosity and grammage) could be dependent on resolution. However, \cite{Hopk20CRtrans} showed that the required diffusion coefficients were not dominated by the cold/dense clouds (since they are small, the residence time of CRs is very small there, even with a lower diffusion coefficient). They showed most of the hadronic $\gamma$ rays comes from volume-filling phases like the WIM and inner CGM. \cite{Bust21crism} also found similar results using high-resolution ISM simulations with CRs. \cite{Chan18cr} found that the $\gamma$ ray emission is much more extended than the cold/thin disks, so the emission is not dominated by the cold gas. Therefore, the effective diffusion coefficient required by observations is likely insensitive to resolution, as long as the WIM and inner CGM are resolved.

Here we only consider averaged quantities over time and radius. However, the dynamical balance of the gas above galactic disks is violated on short time \citep{Beni16diskprebal} and small spatial scales \citep{Gurv20prebal}. Mass fluxes and temperature distributions can also vary due to non-uniform star formation and gravity \citep{Vija20verbal,Kim20SMAUG}. These variations need to be considered in more detailed comparisons with observations.

We do not include AGN feedback in this work (though see \citealt{Su18cr}), which can heat gas and inject CRs (see \S\ref{sec:HIMSFR}), potentially changing properties of the CGM and disk-halo interface. To determine their role, one would have to implement realistic AGN feedback with CR injection in cosmological settings, which is a promising research direction for the future.

Finally, both Hydro+ and CR+ runs are not direct analogues to the MW, but are similar to some actively star-forming $\sim L\star$ galaxies. As seen in Table \ref{tab:SIC}, stellar masses of Hydro+ are higher than the MW's, whereas CR+ galaxies are less massive than the MW. 

Both of them have higher specific star formation rates than the MW's, although they are not atypical in star forming galaxies at these stellar masses, see e.g. \cite{Sali07SFRz0}. In fact, the SFR of MW is a factor of 2-3 lower than the typical $L\star$ galaxies \citep{Licq15MWSFR}. 

Hence, unsurprisingly, some quantities in our simulations deviate from the MW's value. For example, the HI vertical velocity dispersions of our {\bf m12} runs are around 30 km/s (the rightmost points in Fig. \ref{fig:speedmwsfr}), which are three times higher than the MW's value \citep{Boul90} \footnote{ But our {\bf m12} runs still follow the observed $\sigma_{\rm HI}$ - $\Sigma_{\rm SFR}$ relation, because their $\Sigma_{\rm SFR}$ are also higher than the MW's value ($\Sigma_{\rm SFR,MW}\sim 5\times 10^{-3}\;{\rm M_\odot/yr/kpc^2}$; \citealt{Kenn12SFreview}).}. How to reproduce the MW properties (lower velocity dispersions and SFRs) would be one potential topic for the future.

%% file: conclusions.tex
\section{Conclusions}
\label{sec:conclusion}
We study the disk-halo interaction using cosmological simulations of $L\star$ galaxies with the FIRE feedback scheme that includes CR feedback with diffusion and streaming. Our simulations include the full range of processes relevant for the evolution of gas around disks, e.g. galactic fountains, inflows, and outflows. We focus on the pressure balance and the distribution of the multiphase gas near galaxies in simulations with and without CRs. Our major findings are summarized below:

\begin{itemize}
\item The extra-planar gas of $L\star$ galaxies in our sample is in approximate dynamical equilibrium, but not in hydro-static equilibrium. In all of the runs, gas is supported predominately by kinetic pressure near the galactic midplane. Without CRs, significant thermal and kinetic pressure are required to support gas far above ($>5\;{\rm kpc}$) the disks. However, if included, CRs become the major support at a large height above the galactic disk; 

\item Pressure from gas motion (with scale $\gtrsim 1\; {\rm kpc}$) dominates the mid-plane pressure in all of the runs. CR feedback does not affect the relation between SFR surface density and mid-plane turbulent velocity (\S\ref{sec:midturSFR}). 

\item Our simulations without CRs form the traditional ``galactic fountains'' (\S\ref{sec:vergasflow}), where winds are mostly trapped due to the strong gravitational potential, efficient radiative cooling, and hot halo confinement. With CR-dominated halos (for $z>2\;{\rm kpc}$), the disk-halo interface is governed by CR-driven warm winds and hot gas bubbles. Superbubble-driven winds can escape the inner halo more easily with CRs. Fig.\ref{fig:schematic} shows the schematic diagrams of these processes.

\item CRs drastically modify gas phase structures above galactic disks (\S\ref{sec:verticaldensity}). CRs can boost the scale height of the warm-hot gas and suppress the hot gas, so the warm-hot ($T\sim 2\times 10^4-5\times10^5 {\rm K}$) gas becomes the dominant voluming-filling component away from galaxies. 

\item The soft X-ray luminosity of the CR runs is close to the lower bound of the observation. Future higher resolution simulations with a more realistic CR propagation models are needed to clarify this tension.

\end{itemize}

Note that we do not necessarily assert that $L\star$ galaxies have CR-dominated halos (see also \citealt{Ji19CRCGM}). However, several lines of arguments suggest that CRs play an important role in the disk-halo interface. First, CR pressure is strong in the ISM \citep{Boul90,Ferr01}, and should be significant also in the disk-halo interface due to fast diffusion \citep{Chan18cr}\footnote{For example, using the observed CR species, \cite{Trot11} inferred the CR halo in the MW has a height at least around 5 kpc.}. Second, there is evidence of CR-driven winds in moderately star forming galaxies \citep{Hees18obsCRwind,Schm19crwind}. Third, CRs (and other non-thermal pressure) are a possible explanation of the prevalent WHM observations \citep{Ji19CRCGM,Faer17CRCGM}. Therefore, CRs are an indispensable component in future studies of the disk-halo interface.